\documentclass[aip,pop,article-title]{revtex4-2}
\usepackage{amssymb}
\usepackage{amsmath}
\usepackage{graphicx}
\usepackage{hyperref}
\usepackage{bm}
\ifx\affiliation\undefined 
\def\affiliation#1{\date{\normalsize #1}}
\usepackage[margin=1in]{geometry}
\usepackage[sort&compress,numbers]{natbib}
\fi
\def\energy{{\cal E}}
\def\etothe#1{{\rm e}^{#1}}
\def\sech{{\,\rm sech}}

\begin{document}

\title{Ion Hole Equilibrium and Dynamics in One Dimension}
\author{I H Hutchinson}
\affiliation{Plasma Science and Fusion Center,\\ Massachusetts Institute of
  Technology,\\ Cambridge, Massachusetts, 02139, USA}

\ifx\altaffiliation\undefined\maketitle\fi 
\begin{abstract}
  Electrostatic solitary waves with negative potential (ion holes) are
  analyzed theoretically using a generalization of the treatment
  recently developed for slow electron holes. It is shown that an
  often-cited criterion for their existence is mistaken and they can
  in fact exist for a wide range of ion to electron temperature
  ratios. Shifts of the hole velocity $v_h$ relative to the ion
  distributions systematically decrease the permitted hole depths,
  which become extremely small by $v_h/v_{ti}\sim 2$. Ion holes are
  usually unstably accelerated by electron reflection forces which are
  calculated numerically and analytically for the resulting asymmetric
  potential structure. The timescale of this acceleration is
  proportional to the ion plasma period, and generally longer than the
  ion bounce time in the potential well. Thus, ion holes behave like
  approximately rigid entities and even when unstable can survive much
  longer than the typical transit time of a satellite, so as to be observable.
\end{abstract}
\ifx\altaffiliation\undefined\else\maketitle\fi  

\section{Introduction}

A solitary negative electrostatic potential structure sustained in a
plasma by a deficit of trapped ions is called an ion hole. Ion hole
theory springs from \citeauthor{Bernstein1957} (BGK mode) equilibrium
analysis\cite{Bernstein1957}, and from its subsequent development
\cite{Bujarbarua1981,Schamel1982,Dupree1982}, which accompanied ion
hole observations in laboratory experiment
\cite{Pecseli1981,Johnsen1987} and in numerical
simulation\cite{Sakanaka1972,Pecseli1984,Berman1985,Goldman2003}. More
recently, unambiguous identification of the polarity of solitary
structures in space observations has shown that ion holes occur widely
at the Earth's bow shock \cite{Wang2021}, the Plasma
Sheet\cite{Wang2022}, and in the near-Sun solar wind\cite{Mozer2021}
as well as in the auroral region\cite{Bounds1999}. These observations,
together with recent related developments in the theory and
observation of positive potential solitary waves (i.e.\
\emph{electron} holes)\cite{Hutchinson2016,Hutchinson2017}, provides
motivation to revisit ion hole theory.

The mechanisms that are expected to produce ion holes are kinetic ion
streaming instabilities (e.g.\ \cite{Muschietti2008}) including
ion-ion, Buneman, and ion-acoustic instabilities; but the present work
reports on ion holes' equilibria once formed and subsequent dynamics,
rather than on formation mechanisms. The theory depends upon an
assumption that ion holes persist long enough to be treated as
quasi-static Vlasov equilibria, in which case the velocity
distribution function is constant on particle orbits. But whether that
assumption is justified itself depends on the fate of an ion hole
regarded as a composite object accelerating under the influence of the
rest of the plasma.  Although trains of ion holes are often observed,
the present analysis concentrates on solitary holes in a prescribed
background plasma.  The treatment draws heavily on related theory of
electron hole dynamics\cite{Hutchinson2021d,Hutchinson2021c}, but the
large mass difference makes interaction with the opposite charge
species more important for ion holes.

Section \ref{sec1} explains the generalized notation in terms of
attracted and repelled species and how reflection affects the repelled
distribution function. Section \ref{sec2} addresses hole equilibria in
the presence of symmetric repelled species distributions and
calculates the criteria for their existence, correcting an important
prior error in the literature. Section \ref{sec3} shows how to
calculate the momentum balance of a hole when it is subject to
acceleration arising from asymmetric reflection, and gives analytic
values for the acceleration growth rate from initial
equilibrium. Discussion of the significance and limitations of the
present analysis is given in Section \ref{sec4}.

\section{General phase-space hole equilibria}\label{sec1}

Electron holes can, and often do, have speeds much higher than ion
thermal speed, allowing them to be considered unaffected by ion
interactions. When this is not the case they are considered ``slow''.
By contrast, ion holes exist only when the spread of the ion
distribution function $f_i(v)$ in the hole rest frame makes
$f_i(0)\not=0$. The ion velocity spread is much smaller for physical
mass ratios than the electron velocity spread; so ions are thus
practically always slower than electrons and ion holes are always
``slow'' in the sense that they reflect some of the repelled species:
electrons. To treat such ``slow'' situations, and transfer intuition
and calculation from electron holes to ion holes it is convenient to
treat both ions and electrons on an equal footing as follows.

\begin{figure}
  \center
  \includegraphics[width=.8\hsize]{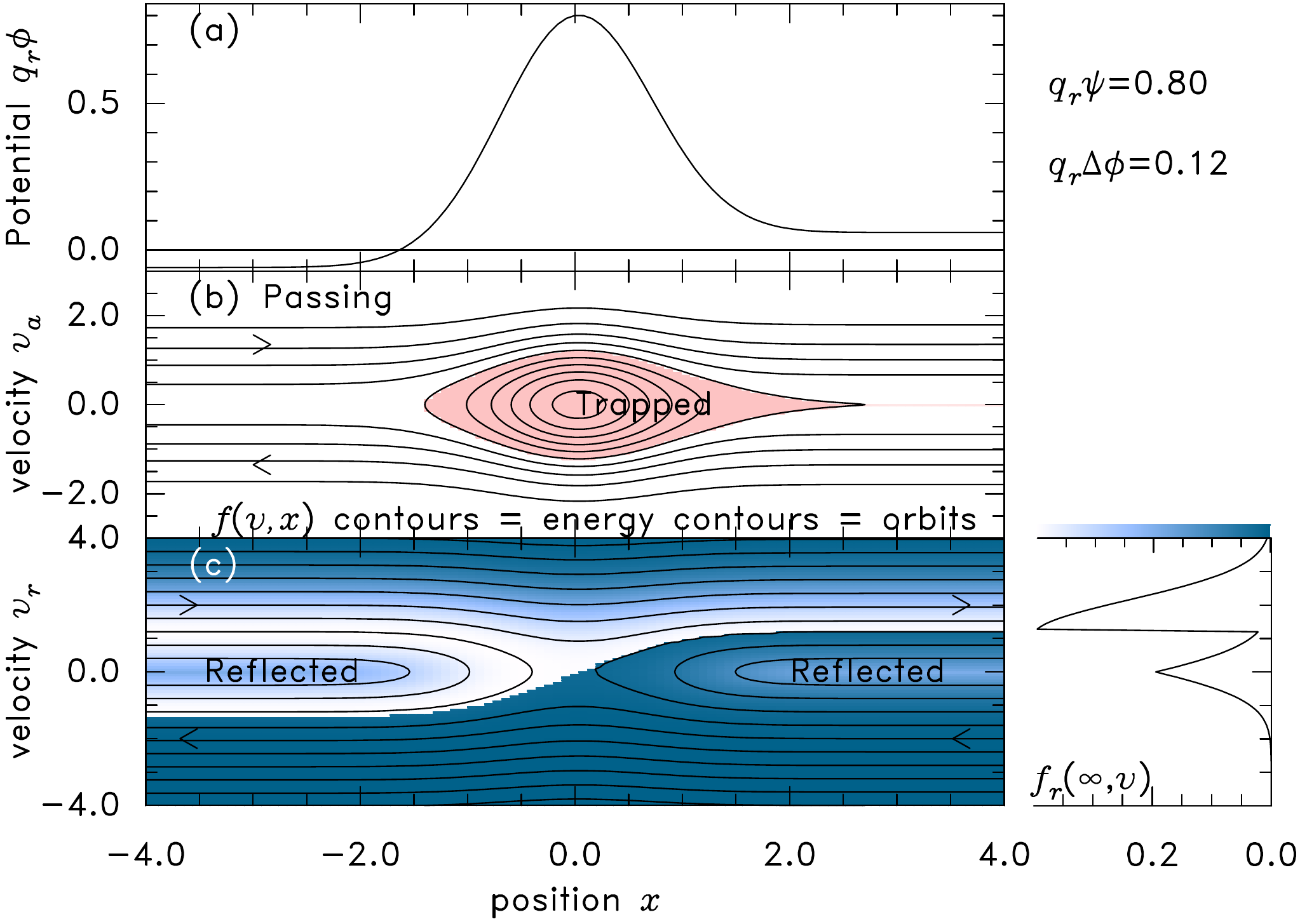}
  \caption{Generic phase-space hole structure. (a) potential, (b)
    attracted species phase space density contours, (c) repelled
    species contours.\label{structure}}
\end{figure}

We consider a local region of potential $\phi(z)$ having only a single
stationary point (${\partial\phi\over \partial z}=0$) at $z=0$, and
zero slope at $|z|\to \infty$. The local region is called the hole, as
illustrated in Fig.\ \ref{structure}(a). The hole is (initially) at
rest ${\partial \phi\over\partial t}=0$ in its frame of
reference. Taking the zero of potential to be
$(\phi(+\infty)+\phi(-\infty))/2=0$, the sign of $\phi(0)$ (which is
necessarily minus the sign of $d^2\phi/dz^2|_0$) attracts particles
toward $z=0$ if their charge is opposite that of $\phi(0)$, or repels
them if it is the same. We will use subscripts $a$ and $r$ to denote
attracted and repelled. So, for an ion hole ($\phi(0)\le 0$),
$i\leftrightarrow a$, and $e\leftrightarrow r$.  We use units that are
non-dimensionalized using the mass of the species under consideration
but the (effective external) temperature $T_a$ of the attracted
species.  Thus, time is measured in units of
$\omega_{p}^{-1}=\sqrt{\epsilon_0 m/n q^2}$, velocity in units of
$\sqrt{T_a/m}=1$, length in units of
$\lambda_{Da}=\sqrt{n_{\infty} q^2/\epsilon_0T_a}$, potential in units
of $|q|T_a$. Quantities without $a$ or $r$ subscripts here correspond
to whichever species is under consideration. For example,
electron and ion velocities are normalized respectively to their
(different) thermal velocities $\sqrt{T_a/m_{e/i}}$. For simplicity we
regard the charge magnitudes $|q|$ to be the same for both species
(singly charged ions), so outside the hole, neutrality gives
$n_{a\infty}q_a^2=n_{r\infty}q_r^2$. The normalized density of both
in the absence of a hole is $n_\infty=1$, and the normalized magnitude
of charge is $|q|=1$. The units of time (and velocity) are different
for different species by the square root of the mass ratio. It is
convenient to introduce an abbreviation for the normalized potential
energy of the repelled species: $\varphi \equiv q_r\phi$ which is
always positive at $z=0$. We denote the maximum repelled potential
energy as $\psi$, always positive.  We allow for the possibility that
there is a potential difference
$\Delta \varphi =\varphi(+\infty)-\varphi(-\infty)$ across the hole,
but not that the difference is so great that it eliminates the local
maximum of $\varphi$.

Provided the transit time of the particles is short compared with any
hole potential changes (so $\phi$ is effectively time-independent in the hole
frame), the repelled species' velocity distribution is determined by
the constancy of $f_r$ on orbits (Vlasov equation), the conservation
of energy ($\energy_r=v^2/2+q_r\phi=v^2/2+\varphi $), and the value of
$f_r(v)=f_r(\sqrt{2[\energy_r-\varphi]})$ at $|z|=\infty$. The
velocity distribution of only the \emph{untrapped} particles of the
attracted species is determined by distant conditions; those particles
have $\energy_a=v^2/2-\varphi \ge -|\Delta\varphi/2|$. The attracted species particles
with $\energy_a< -|\Delta\varphi/2|$ are trapped; their orbits are localized in the
hole; and the distribution on them is determined by the initial
conditions of formation of the hole. Fig.\ \ref{structure} (b) and
(c), illustrate the attracted and repelled species' phase-space distribution
contours for a case where the incoming distributions are Maxwellians
of the same temperature but the repelled velocity distribution is
shifted (in the hole frame) by a normalized velocity of
$\bar v_r=1.2$. This shift causes discontinuities in $f_r$ at the boundary
between passing and reflected orbits. The resulting ($z\to\infty$)
distribution is illustrated in the side panel, which indicates the
intensity calibration for the repelled species contours.

It is conceptually and numerically advantageous to regard the
attracted species as consisting of a reference distribution given by
the passing particles but including a flat distribution (independent
of $v$) of trapped particles, matching the untrapped distribution at
their join where total energy in the hole frame, $\energy$, is
$-|\Delta\varphi/2|$. The trapped region is pink in Fig.\
\ref{structure}(b). This ``flat-trapped'' distribution $f_f(v)$
governs the attracted species when there is no ``hole'' in the
distribution. It is what would arise if the hole potential were
(hypothetically) grown infinitessimally slowly. By itself, it is
generally unable to sustain the potential self-consistently through
Poisson's equation, and requires an additional contribution on trapped
orbits only, $\tilde f$, which is negative, and represents the deficit
of the phase-space density.  $\tilde f$ is responsible for the
required charge density with sign the same as $\phi$ near the hole
center. The corresponding particle densities $n=\int f dv$, are
functions of $\phi$: repelled $n_r$, and attracted $n_f+\tilde n$.

Poisson's equation for the hole is then
\begin{equation}
  \label{eq:Poisson}
  {1\over q_a}{d^2\phi\over dz^2}=-{d^2\varphi \over dz^2} = n_r-n_f-\tilde n.
\end{equation}
If we regard $\varphi(z)$, and $f_a$, $f_r$ at $|z|=\infty$, as prescribed,
then all terms except $\tilde n$ are prescribed, and $\tilde n$ is
determined. Then $\tilde n =\int \tilde f dv$ is an integral equation
determining the trapped distribution deficit $\tilde f$.  It is Abel's integral
equation whose known solution (on the higher $\varphi(\pm\infty)$ side) is
\begin{equation}
  \label{eq:abel}
  \tilde f(v_0) = {1\over \sqrt{2}\pi}\int_{|\Delta\varphi/2|}^{{\psi}-v_0^2/2}
    {d\tilde n\over d\varphi }{d\varphi \over
      \sqrt{{\psi}-v_0^2/2-\varphi } }
    ={1\over \sqrt{2}\pi}\int_{|\Delta\varphi/2|}^{-\energy}
    {d\tilde n\over d\varphi }{d\varphi \over
      \sqrt{-\energy-\varphi } },
\end{equation}
where $v_0$ denotes the attracted species velocity at the center of
the hole $z=0$ ($\varphi=\psi$), which is $v_0=\sqrt{2(\energy+{\psi})}$.

\section{Equilibria with Maxwellian Background}\label{sec2}

Suppose the repelled species velocity distribution $f_r(v)$ is an
unshifted Maxwellian in the frame of the hole. The repelled species density
is then $n_r=\etothe{-\varphi /\theta}$, where $\theta=T_r/T_a$.
Take the untrapped attracted velocity distribution to be a Maxwellian
distribution with a velocity shift $\bar v_a$ in the hole frame. Then the
flat trapped distribution at potential $\varphi$ is
\begin{equation}
  \label{eq:flattrap}
  f_f(v,\varphi)
  ={1\over \sqrt{2\pi}}\etothe
  {-(\sigma_v\sqrt{2\,{\rm max}(\energy_a,0)}-\bar v_a)^2/2}
  ={1\over \sqrt{2\pi}}\etothe
  {-(\sigma_v\sqrt{{\rm max}(v^2-2\varphi ,0)}-\bar v_a)^2/2}
\end{equation}
where $\sigma_v=v/|v|$ is the sign of $v$, and all velocities are
expressed in the hole rest frame. The integral over all velocities
gives the density $n_f(\varphi)=\int f_f(v,\varphi)dv$.

\subsection{Unshifted Maxwellian attracted species}
It is easy to show\cite{Hutchinson2017}
that when $\bar v_a=0$, the resulting flat-trapped density is
\begin{equation}
  \label{eq:nft}
  n_f=  {2\over \sqrt{\pi}}\varphi ^{1/2}+
  \etothe{\varphi }{\rm erfc}(\varphi ^{1/2}).
\end{equation}
Poisson's equation for the hole is then
\begin{equation}
  \label{eq:Poisson2}
  -{d^2\varphi \over dz^2} = n_r-n_f-\tilde n
  = \etothe{-\varphi /\theta}- [{2\over \sqrt{\pi}}\varphi ^{1/2}+
  \etothe{\varphi }{\rm erfc}(\varphi ^{1/2})] -\tilde n.
\end{equation}
Denoting the (prescribed) field divergence as
$n_\varphi\equiv{d^2\varphi \over dz^2}$. We have
$\tilde n= n_\varphi+n_r-n_f$ which is to be substituted into eq.\
(\ref{eq:abel}). Since that equation is linear in the density
deficit, we can treat the three density terms separately and then add
them up. The repelled species derivative needed for eq.\ (\ref{eq:abel})  is
${dn_r\over d\varphi }= -\etothe{-\varphi /\theta}/\theta$, and the
flat-trapped contribution is
${dn_f\over d\varphi }= \etothe{\varphi }{\rm erfc}(\varphi ^{1/2})$.

As an illustrative choice let us suppose that $\Delta \varphi=0$ and the
hole potential is of the form
\begin{equation}
  \label{eq:potl}
  \varphi(z) = \psi \sech^\ell\left({z\over\ell\lambda}\right).
\end{equation}
The parameter $\ell$ is usually taken to be 4, but we retain
slightly more generality, because we can still derive closed analytic
expressions. That mathematical convenience arises from the identity
${d^2\over dx^2} \sech^\ell x = \ell^2\sech^\ell x
-\ell(\ell+1)\sech^{\ell+2}x$, from which we find
\begin{equation}
  \label{eq:diverg}
  n_\varphi={d^2\varphi \over dz^2}={1\over \lambda^2}\left[
    \varphi  -{\ell+1\over \ell}{\varphi ^{(\ell+2)/\ell}\over {\psi}^{2/\ell}}
  \right];
\end{equation}
and so
\begin{equation}
  \label{eq:derivdiv}
  {d n_\varphi\over d\varphi }={1\over \lambda^2}\left[
    1 -{\ell+1\over \ell}{\ell+2\over \ell}\left|\varphi\over \psi\right|^{2/\ell}
  \right].
\end{equation}

To avoid divergent slope of $\tilde f$ at $\energy=0$, it is necessary
that $\left.d\tilde n\over d\varphi\right|_{\varphi=0}=0$. That condition 
immediately becomes
\begin{equation}
  \label{eq:shielding}
  {1\over\lambda^2}=\left.{dn_\varphi\over d\varphi}\right|_{\varphi=0}=
\left[-{dn_r\over d\varphi }+{dn_f\over d\varphi
  }\right]_{\varphi=0}={1\over\theta}+1
\end{equation}
The parameter $\lambda$ is precisely the
potential decay length combining repelled and attracted species
screening in attracted Debye length units, because
$\lambda_{Dr}^2=\lambda_{Da}^2\theta$.

\begin{figure}[htp]
  \includegraphics[width=.5\hsize]{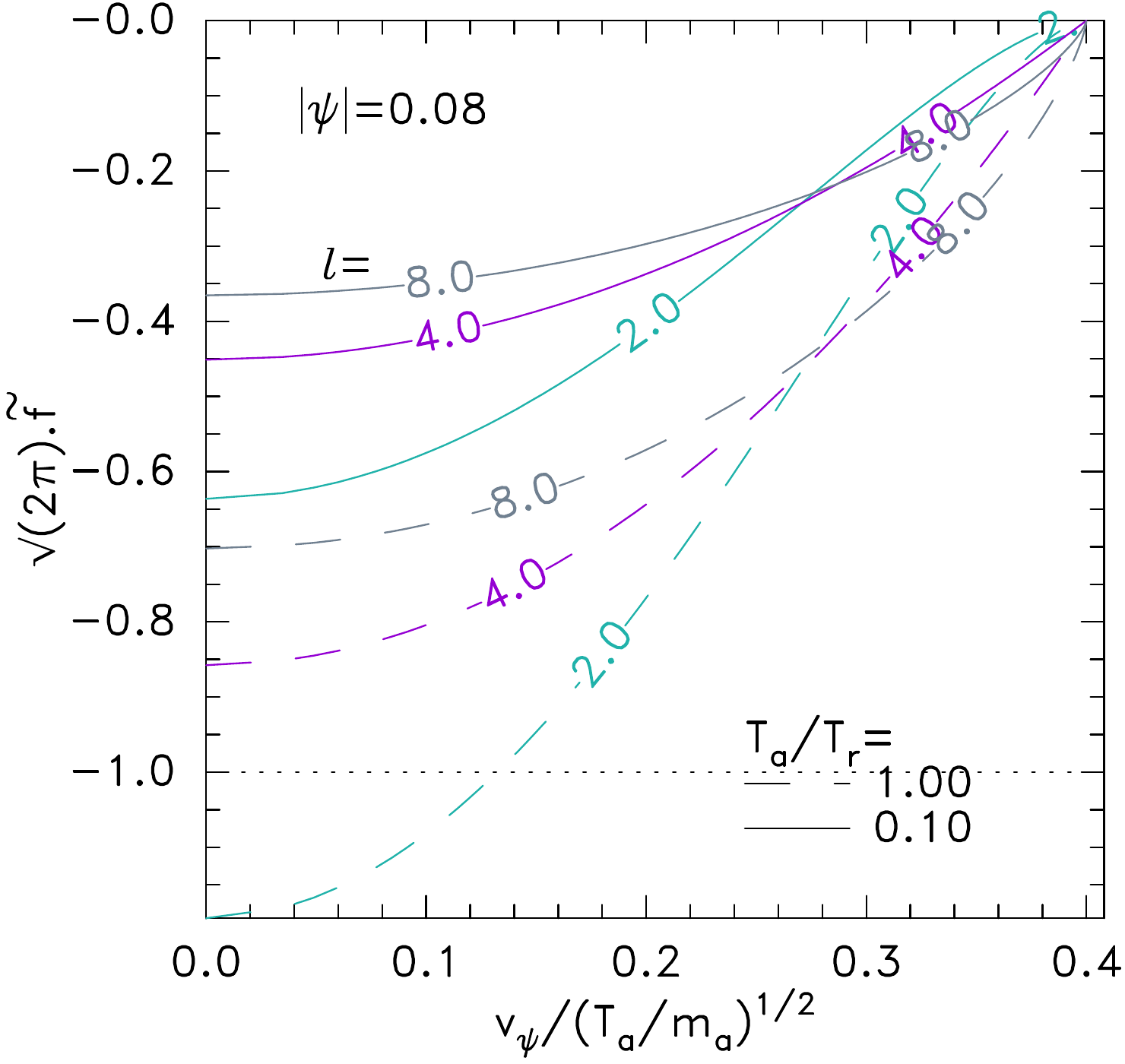}\hskip-1.5em(a)
  \includegraphics[width=.5\hsize]{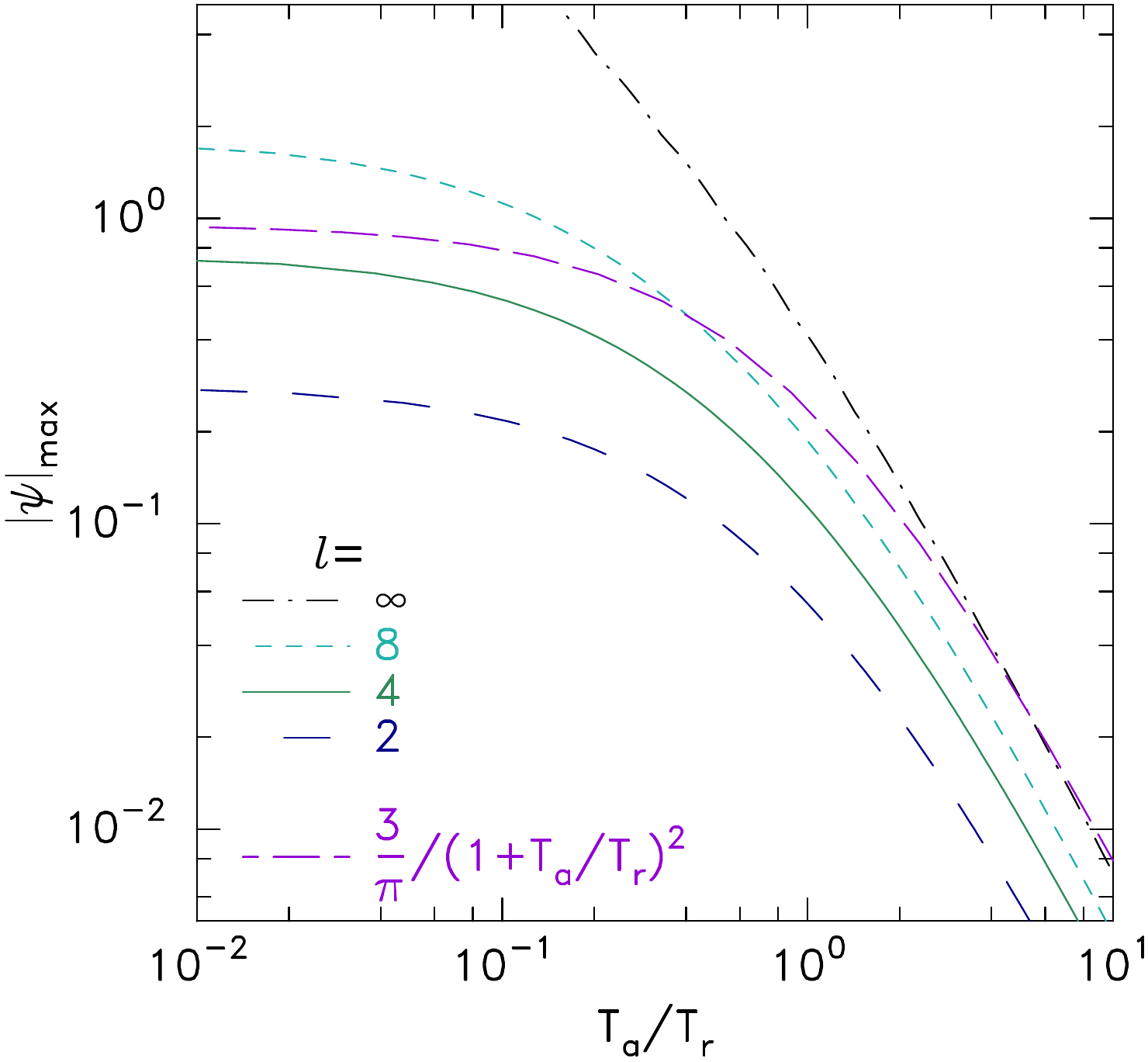}\hskip-1.5em(b)
  \caption{Examples of ion (or electron) holes with unshifted
    background Maxwellian distributions of both species. (a) the
    required trapped deficit at $z=0$, $\tilde f(v_\psi)$, to produce
    a hole of the form $\varphi=\psi\sech^l(z/l\lambda)$ with $l=$ 2, 4,
    or 8, for two ratios of the attracted to repelled species ratio
    ($T_a/T_r$). (b) The maximum ${\psi}$ value
    permitted by non-negativity, as a function of temperature ratio,
    showing a wide allowable range of $T_a/T_r$(=$1/\theta$) at low
    ${\psi}$. The comparison curve ${3\over \pi}/(1+T_a/T_r)^2$ represents an analytic
    estimate (to first order in $\psi$) for a waterbag
    deficit profile: the least constraining of the models described in
    Appendix \ref{appa}.    
    \label{thetaminplot}}
\end{figure}

Substituting the three contributions $dn_r\over d\varphi $,
$-{dn_f\over d\varphi }$, and ${dn_\varphi\over d\varphi }$ into eq.\
(\ref{eq:abel}) we can find three corresponding contributions to
$\tilde f=\tilde f_{r}+\tilde f_f+\tilde f_\varphi$. After considerable
integration effort they can be expressed in terms of the complex
Faddeeva function\footnote{The Faddeeva function is related to the
  Plasma Dispersion function via $Z(z)=i\sqrt{\pi} w(z)$ and to the
  Dawson integral function $F(z)$ by $w(iz)=2F(z)$. Its derivative is
  $w'(z)=2i/\sqrt{\pi}-2zw(z)$; so $-Z'(z)/2=1+i\sqrt{\pi}zw(z)$.}
$w(z)\equiv\etothe{-z^2}{\rm erfc}(-iz)$ , and the Gamma function
$\Gamma(z)$, as follows:
\begin{equation}
  \label{eq:fie}
  \tilde f_{r}=-{1\over \sqrt{2\pi}}\Im[\theta^{-1/2}w(\sqrt{|\energy|/\theta})],
\end{equation}
\begin{equation}
  \label{eq:ff}
  \tilde f_{f}={1\over \sqrt{2\pi}}\Re[1-w(i\sqrt{|\energy|})],
\end{equation}
\begin{equation}
  \label{eq:fphi}
  \tilde f_\varphi=-{\sqrt{|\energy|}\over \sqrt{2\pi}\lambda^2}
  \left[{2\over \sqrt{\pi}} -
    {\ell+1\over \ell}{\ell+2\over \ell}
    {\Gamma(1+2/\ell)\over \Gamma(3/2+2/\ell)}
    \left|\energy\over \psi\right|^{2/\ell}
    \right].
\end{equation}
Eqs. (\ref{eq:fie}) and (\ref{eq:ff}) are equivalent to terms found by
Chen et al\cite{Chen2004}\footnote{Those authors used a Gaussian potential shape so
  their $\tilde f_\varphi$ is different and does not satisfy eq.\
  (\ref{eq:shielding})}. By virtue of the condition
(\ref{eq:shielding}), the coefficient of $\sqrt{|\energy|}$ in an
expansion of $\tilde f$ near $\energy=0$ cancels to zero between the
three terms. However, the term $\propto |\energy|^{1/2+2/\ell}$ (from
$\tilde f_\varphi$) remains. If $\ell>4$, its contribution causes
$f'=d\tilde f/dv$ to diverge at the phase-space separatrix, i.e. as
$|\energy|\to 0$, which is mildly unphysical; but it avoids
non-monotonic behavior which makes some other $\varphi$ shape choices
even less plausible. The standard value $\ell=4$ makes the bracket
$[2/\sqrt{\pi} -(15\sqrt{\pi}/16)|\energy/\psi|^{1/2}]$, and
$\tilde f'(\energy\to0-)$ is finite. For $l<4$,
$f'(\energy\to0-)=0$. These results make no approximations, but assume
that $f(v)$ is a function only of $\energy$, i.e.\ satisfies the
Vlasov equation in a steady potential. They are illustrated in Fig.\
\ref{thetaminplot}(a).

\subsection{Shifted Maxwellian Attracted Species}\label{erroneous}

If the attracted Maxwellian is shifted by $\bar v_a\not=0$ in the frame of the
hole, then there exists no useful closed form expression for
$n_f$ at arbitrary $\varphi$. Numerical curves of $n_f$
have been given elsewhere\cite{Hutchinson2017}. They decrease for non-zero $\varphi$ as $\bar v_a$
increases. Their slope at $\varphi=0$ can be shown by expansion\cite{Schamel1982} to be
${dn_f\over d\varphi }=-{1 \over 2} \Re [Z'(\bar v_a/\sqrt{2})]=\Re[1+i\sqrt{\pi}zw(\bar v_a/\sqrt{2})]$, which is what
determines the attracted species distant shielding length.  Therefore eq.\
(\ref{eq:shielding}) is modified to become
\begin{equation}
  \label{eq:modshield}
0={1\over \lambda^2}- {1\over \theta} + {1 \over 2} Z'_r(\bar v_a/\sqrt{2}).  
\end{equation}

This condition at $\varphi\to0$ (or its equivalent\footnote{Schamel et al
  proceed instead by the differential approach, specifying a negative
  temperature Maxwellian for trapped particles. They find by expansion
  at small $\psi$ that $\varphi=\psi\sech^4(z/4\lambda)$, with
  $\lambda^{-2}=16b\sqrt{{\psi}}/15$ where $b$ depends on the trapped
  temperature and $\bar v_a$, and that (\ref{eq:modshield})
  $1/\lambda^2-1/\theta+{1 \over 2} Z'_r(\bar v_a/\sqrt{2})=0$, which they
  call the nonlinear dispersion relation. }) has been referred to, in
many publications\cite{Schamel1979} based on prescribing
$\tilde f(\energy)+f_f(\energy)$ rather than $\varphi(z)$, as the
``nonlinear dispersion relation''. It is then supposed that it relates
the hole's potential amplitude, speed (relative to the attracted
species), and trapped particle temperature. That is a misleading
perspective, whether for electron holes or ion holes, because it
adopts a particular form for the entire trapped distribution (negative
temperature Maxwellian), which strongly constrains the potential
shape. Actually the relative shape of the hole can have a whole range
of widths upward from a minimum determined by non-negativity of $f$
and the value of ${\psi}$. Avoiding divergent $\tilde f$-gradients
requires primarily that $\varphi$ decays at large distances (small
$\varphi$) with second derivative length scale ($\lambda$) satisfying
eq.\ (\ref{eq:modshield}); and the trapped deficit $\tilde f$ must be
consistent with the rest of the potential shape.

A hole's speed $|\bar v_a|$ relative to the attracted background must not
be so great that the combined particle terms at $|z|\to \infty$ (small
$\varphi$), ${dn_{r}\over d\varphi }-{dn_{f}\over d\varphi }$
(=$-1/\theta+{1\over2}Z'_r$), become positive, otherwise
(\ref{eq:modshield}) cannot be satisfied for real $\lambda$. That
requires $-Z'_r(\bar v_a/\sqrt{2})/2>-1/\theta$. Speed must also be small
enough that the background attracted species distribution at the hole
speed is non-zero, otherwise there is no phase-space density there, in
which it could be a hole. The hole amplitude ${\psi}$ is limited only
by the non-negativity of $f(v)$, which we shall treat in a moment.

A misunderstanding related to eq.\ (\ref{eq:modshield}) is the
belief\cite{Schamel1980}, often
repeated\cite{Bujarbarua1981,Schamel1982,Hudson1983,Pecseli1984,Johnsen1987,Buchanan1993,Griessmeier2002,Eliasson2006a,Schamel2018,Wang2021},
that for an ion hole to exist (``a solution of the nonlinear
dispersion relation [to] exist[s]'') the species' temperature ratio
must satisfy
$T_r/T_a=T_e/T_i=\theta > 1/|{\rm min}(-Z'_r(\bar
v_a/\sqrt{2})/2)|\simeq 3.5$. The minimum of $-Z'_r/2$ over all real
arguments is -0.285 which is where the value 3.5(=$1/0.285$) comes
from. But in fact there is a sign error in this published
criterion. At the $-Z'_r/2$ minimum eq.\ (\ref{eq:modshield}) requires
$-0.285>-1/\theta$ for $\lambda$ to be real.\footnote{A hole requires
  the inequality to be the opposite of the criterion for existence of
  what {\citet{Stix1962}} [section 9.14, equation 71] calls the
  ``zero-damped ion acoustic wave'', because the sinusoidal shape of a
  wave makes ${1\over \lambda^2} \equiv {1\over \phi}{d^2\phi\over dz^2}$
  negative, not positive.}  That is $\theta < 3.5$, not
$\theta > 3.5$.

In any case, the minimum of $-Z'_r/2$ occurs at a speed
$\bar v_a= 2.13$; so this corrected criterion is relevant only for a
hole on the tail of the attracted species' velocity distribution, and
not at all to more typical holes lying deeper within the bulk
population, where $-Z'_r/2$ is positive. The important effect of
finite response of the repelled species is instead that it increases
the required trapped species deficit $|\tilde f|$, and makes the
requirement of non-negativity more stringent. Extensive calculations
of ion hole equilibria have been given by \citet{Chen2004},
who note that they are counter-examples to the erroneous criterion.
Moreover recent observations by \citet{Wang2022} show that ion holes
in the Plasma Sheet are not subject to the erroneous criterion, since
they occur where $T_e/T_i\lesssim 0.3$.

Detailed calculations of hole structure with $\sech^l$ potential shape
and unshifted distributions, illustrating the actual hole feasibility,
are given in Fig.\ \ref{thetaminplot}.
The mathematics (equations \ref{eq:fie} to \ref{eq:fphi}) is identical
for ion and electron holes when expressed in our present normalized
units. Non-negativity of the trapped distribution requires that
$\tilde f > -f_a(\energy=0)$, which is $\sqrt{2\pi}\tilde f > -1$ for
an unshifted Maxwellian, marked with a dotted line in Fig.\
\ref{thetaminplot}(a). Thus, of the six cases shown there, only the
$l=2$ case with equal temperatures $T_a/T_r=1/\theta=1$ is impossible. All the
others are possible equilibria. If $\psi$ were decreased, none would
be impossible; if it were increased, more cases would require
unphysical negative $f_a$.

Fig.\ \ref{thetaminplot}(b) shows the maximum allowable hole amplitude
${\psi}_{max}$ as a function of $T_a/T_r=1/\theta$, for several values
of $\ell$. That maximum occurs when the trapped distribution
$f_f+\tilde f=f(0)+\tilde f=f(0)+\tilde f_r+\tilde f_f+\tilde
f_\varphi$ reaches zero at $z=v=0$ (i.e.\ at $\energy=-{\psi}$). The
root of this expression is found numerically. In addition, to show
that the contradiction of the widely cited criterion is not somehow
attributable to the fact that the ``integral equation'' method is used
to obtain it, Fig.\ \ref{thetaminplot}(b) also includes the upper
boundary of ${\psi}$ allowed by non-negativity, obtained analytically
using the ``differential equation'' method for a prescribed
$f_f+\tilde f$ shape. The $\tilde f$ shape for this curve is two-level
(waterbag), which of all the monotonic ``power deficit''
models\cite{Hutchinson2021a}, allows the highest ${\psi}_{max}$. See
Appendix \ref{appa} for details. Since the resulting expression,
${\psi}_{max}={3\over \pi}/(1+T_a/T_r)^2$, approximates the $n_f-1$
and $n_r-1$ as linear in $\varphi$, it is accurate only to first order
in ${\psi}$. When ${\psi}_{max}$ is small, because $T_a/T_r\gtrsim1$,
it agrees remarkably well with the $\sech^l$ model, especially for
large $l$, which is the potential shape that is least
constraining. The curve for a Gaussian potential shape appearing in
\citet{Chen2004}[Fig 1b] is very similar to the $l=\infty$ curve,
both of which might over-estimate the maximum allowed $\psi$ at
$T_a/T_r\lesssim 1$, because their divergent gradient at $\energy=0$ is
unphysical.

\begin{figure}
  \includegraphics[width=0.5\hsize]{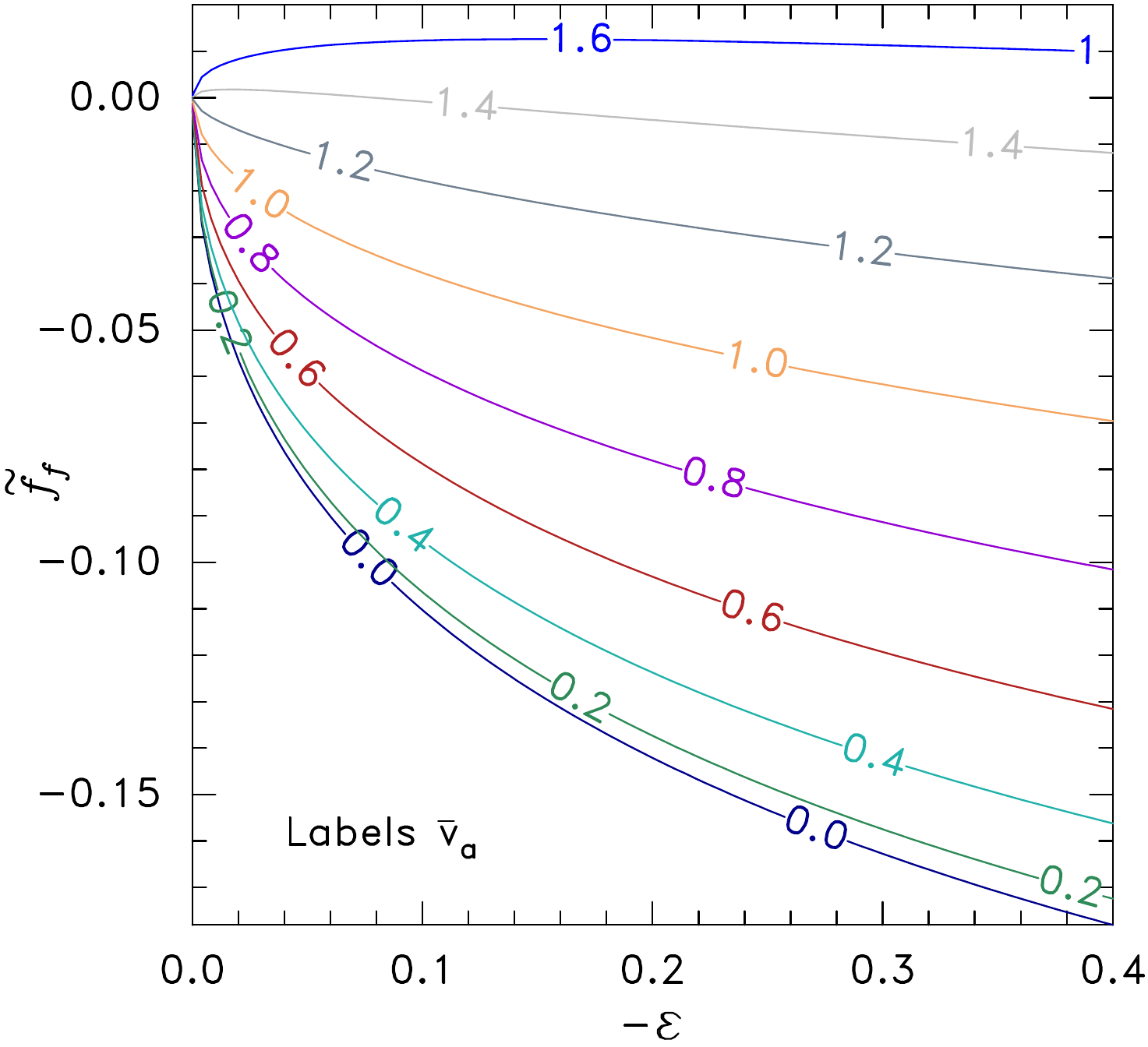}\hskip -2em(a)\hskip1em 
    \includegraphics[width=0.48\hsize]{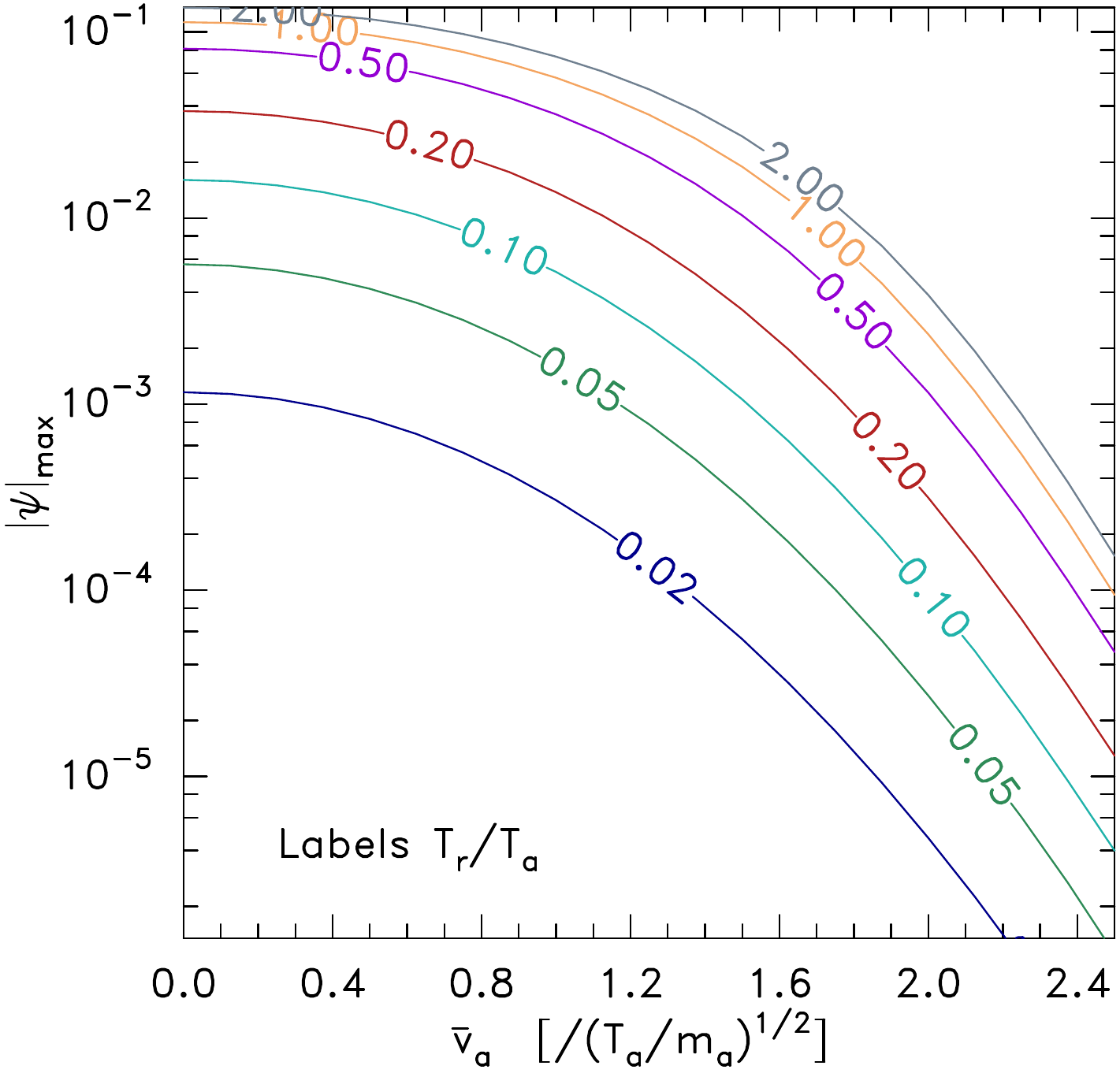}\hskip -2em(b)\ 
    \caption{(a) The contribution arising from the flat-trapped
      distribution to $\tilde f(\energy)$: the required trapped
      distribution deficit as a function of energy. Line labels
      indicate Maxwellian shift $\bar v_a$. (b) The hole potential
      amplitude permitted by non-negativity of $f$ as a function of
      attracted Maxwellian velocity shift. The potential
      profile uses $l=4$. Line labels indicate
      temperature ratio $T_r/T_a=\theta$.\label{fig:ftildef}}
\end{figure}
A background distribution velocity shift $\bar v_a$ of the attracted
species relative to the hole, induces no hole asymmetry, but has two
effects. One is a reduction in its flat-trapped density response to
the hole potential ${dn_f\over d\varphi }$ and thus the magnitude of
the contribution $\tilde f_f$ to the required trapped depletion. In
figure \ref{fig:ftildef}(a), this variation is plotted.  The shift
$\bar v_a$ of the attracted species does not affect $\tilde f_r$, and
since it acts in the same direction at low $\bar v_a$ as $\tilde f_f$,
the $\tilde f_r$ comes to dominate the particle $\tilde f$
contributions as $\bar v_a$ increases toward unity.  The other effect
of increasing $\bar v_a$ is to reduce the level of the flat trapped
region to $f(0)={1\over \sqrt{2\pi}}\exp(-\bar v_a^2/2)$. As
$\bar v_a$ is increased, therefore, the hole amplitude must be reduced
to avoid the need for trapped deficit corresponding to an unphysical
negative trapped distribution function. Fig.\ \ref{fig:ftildef}(b)
shows the solutions of the equation
$\tilde f(\energy=-{\psi})+f(\energy=0)=0$ (i.e. the total
$f_a(\energy=-\psi)=0$) using equations (\ref{eq:fie}), and
(\ref{eq:fphi}) together with the numerical evaluation of
$\tilde f_f$. The result is ${\psi}_{max}$, the maximum value of
${\psi}$ permitted by non-negativity of the (monotonic) total trapped
distribution function.  It shows that large $\bar v_a$ permits only
exponentially small potentials, because there is so small an attracted
background $f(0)$ in which to have a hole. Lower temperature ratio
$T_r/T_a$ gives a lower ${\psi}_{max}$ because the repelled species
response is stronger, requiring enhanced deficit $|\tilde f|$ to
overcome its influence.

\section{Hole dynamics and stability}\label{sec3}

In this section it is more convenient to speak in terms of velocity
relative to some fixed frame, set by the particle velocity
distributions, not the hole. In this frame the hole has a speed
$v_h$. Electron hole speeds are often large compared with the thermal
speed of the ions (the reflected species). In that case, repelled
species response can be ignored.

When the repelled species response cannot be ignored however, because
hole velocity is comparable to typical repelled velocity, repelled
velocity distributions that are asymmetric (about the hole velocity)
give rise to net reflection force on the hole, which causes the hole
to accelerate. Such holes moving relative to a single shifted
Maxwellian repelled species background distribution are thus not
steady equilibria. Instead, the time dependence increases their
effective velocity shift. Moreover holes with zero velocity relative
to a Maxwellian repelled background, while they are (by symmetry)
equilibria, are generally unstable to growing acceleration. This
reflection mechanism was proposed as a possible cause of ion hole
growth in early theory by Dupree\cite{Dupree1982,Dupree1983} who
analyzed the dynamics of holes treated as composite objects. There is
good simulation evidence for the ``self-acceleration'' instability of
electron holes\cite{Eliasson2004,Eliasson2006}. Analytic theory of
electron hole self-acceleration by interaction with ions agrees with
simulations\cite{Zhou2016}.

Because velocity spread of ions is far smaller than electrons, ion
holes essentially never move fast enough relative to the electron mean
velocity to be able to ignore electron reflection.  Therefore an
important theoretical question arises as to whether and for how long
ion holes, for which the repelled species is always significant, can
exist. Acceleration of the hole velocity relative to attracted
background (i.e.\ $\dot v_h=-\dot{\bar{v_a}}$) causes evolution of the hole
amplitude ${\psi}$, both because the external distribution at $f(v_h)$
changes, and because the amplitude must remain below the relevant
${\psi}_{max}$ curve of Fig.\ \ref{fig:ftildef}(b). So once $|\bar v_a|$,
the hole's speed relative to a Maxwellian attracted species' mean
velocity, exceeds $\sim 2(\times v_{ta})$ its amplitude becomes
negligible. Unlike electron holes, this dissipation happens for ion
holes long before their interaction with the repelled distribution
ceases. The growth of the self-acceleration instability to speed
$\gtrsim 2$ is thus one important limit on the lifetime of an ion
hole.

\subsection{Asymmetric Repelled Distributions}

\begin{figure}[htp]
\centering
  \includegraphics[width=0.6\hsize]{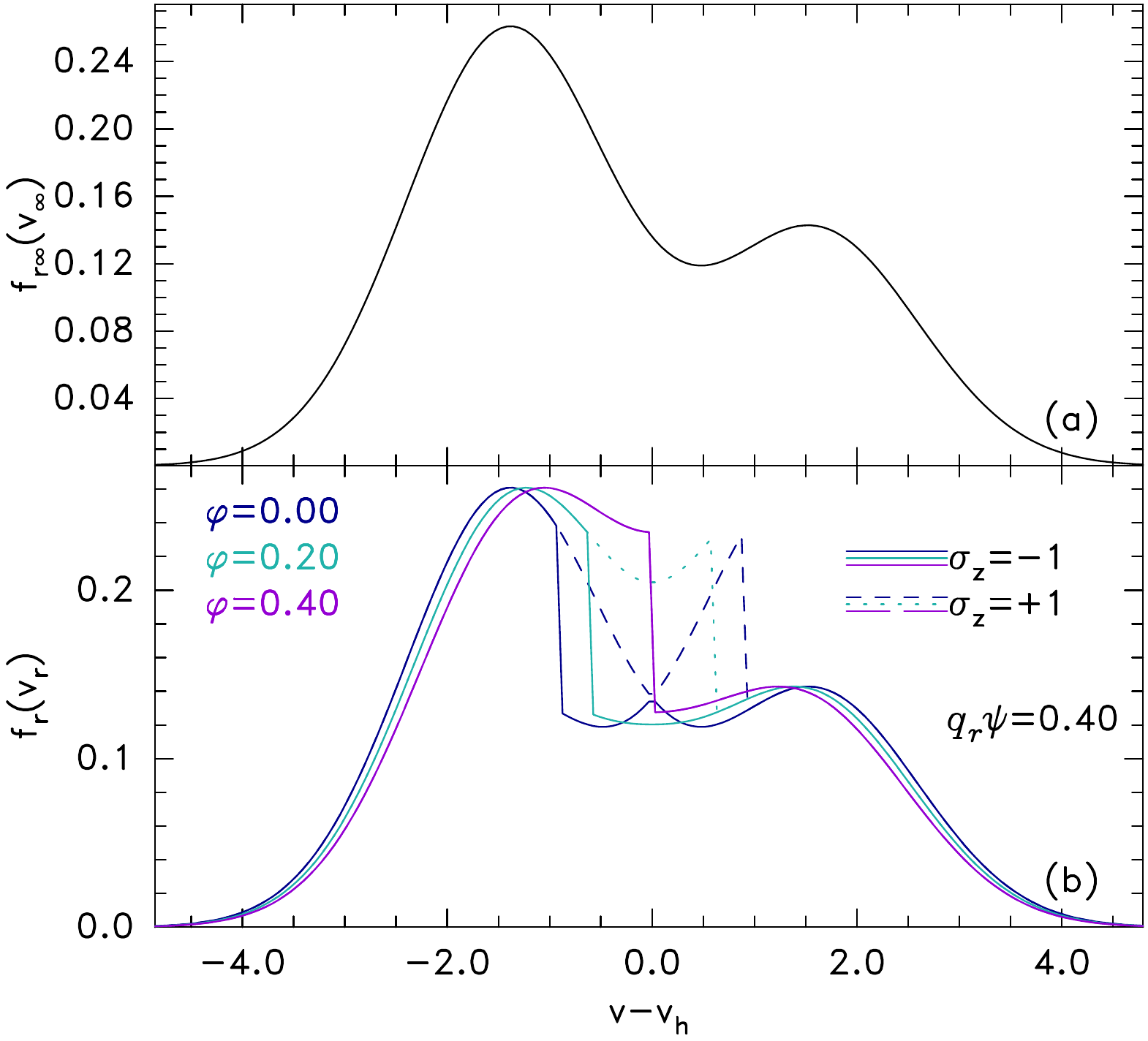}
  \caption{(a) An example background repelled species
    distribution. (b) The corresponding local distribution functions at
    selected potentials $\varphi$ on either side ($\sigma_z=\pm1$) of the
    potential peak (where $\varphi=\psi$). \label{fvofvplot}}
\end{figure}

Asymmetric repelled species distribution (relative to the hole
velocity $v_h$) causes a difference in the distant repelled species
density on either side of the hole.  That requires a potential
difference between the distant potentials on either side of the hole
$\Delta\varphi\not=0$, to satisfy distant
neutrality\cite{Hutchinson2021d}. The mechanism is illustrated in
Fig.\ \ref{fvofvplot}, which can be considered to show vertical slices
through the contours of Fig.\ \ref{structure}, except that a more
complicated background repelled distribution $f_{r\infty}(v_\infty)$ shown in the top panel
Fig.\ \ref{fvofvplot}(a) is considered. It represents the value of
$f(v)$ on incoming orbits (${\rm sign}(v)=-\sigma_z$) at
$|z|=\infty$. Fig.\ \ref{fvofvplot}(b) shows local distributions
$f_r(z,v_r)$ at different potentials on either side of the hole
potential peak (${\rm sign(z)}=\sigma_z=\pm1$).
Reflected orbits for $\sigma_z=-1$ (having negative velocity $v-v_h$)
are populated by incoming orbits at $z=-\infty$ with positive
velocity, while for $\sigma_z=+1$ reflected orbits with positive
velocity are populated by negative velocity orbits at
$z=\infty$. Consequently there is a $f_r(v_r)$ discontinuity at the
energy equal to the potential peak $\psi$, which is at
${1\over2}(v_r-v_h)^2=\psi-\varphi$. As the position $z$ moves inward
from $\pm\infty$ and the potential $\varphi$ climbs, the discontinuity velocity moves
toward $v_h$ and exactly reaches it when $\varphi=\psi$. There are
no $z$ positions with $\varphi>\psi$. 

\begin{figure}[htp]\centering
  \includegraphics[width=.7\hsize]{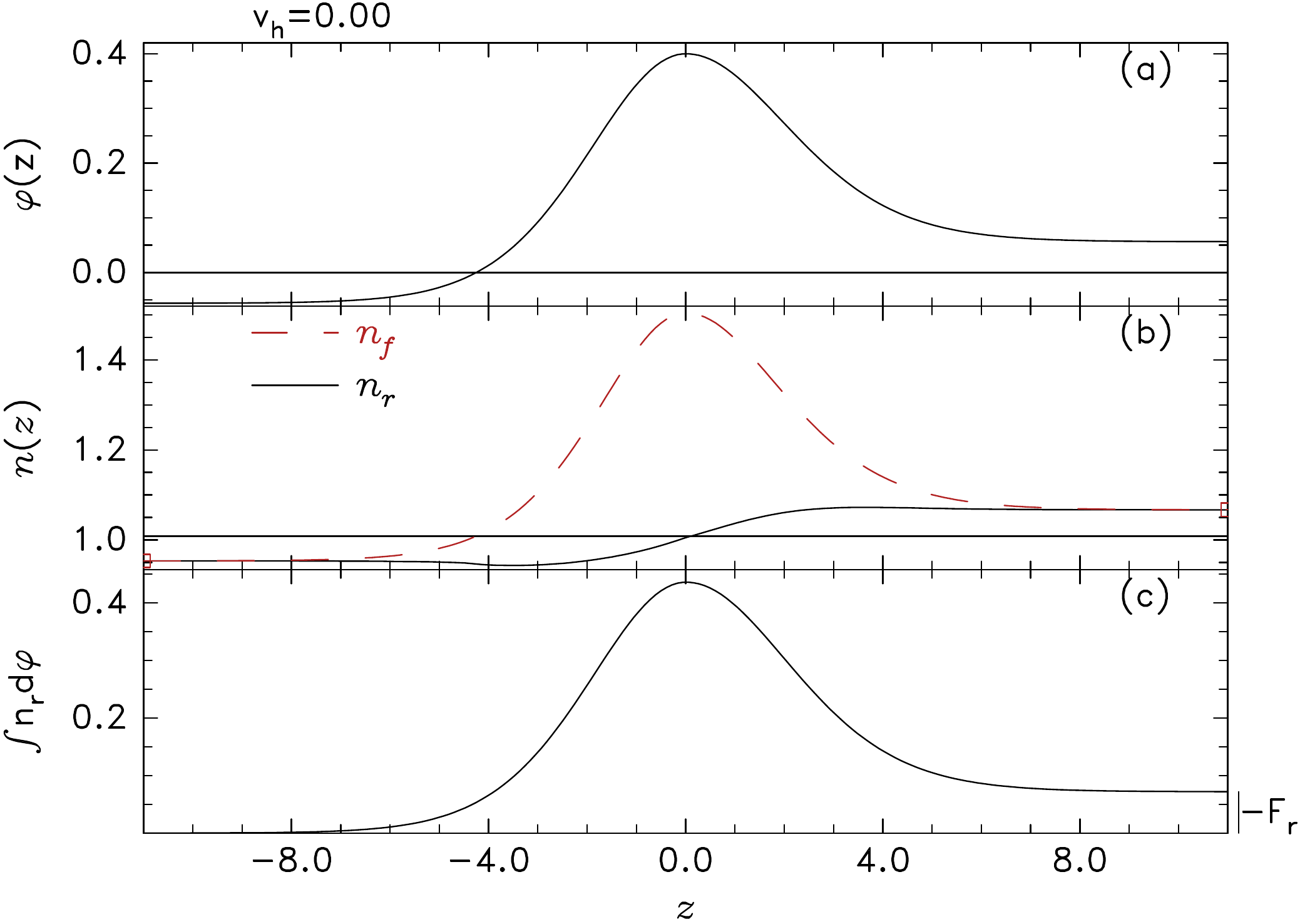}
  \caption{Spatial dependence of (a) potential, (b) density, and (c)
    cumulative force $\int_\infty^z n_rd\varphi$ on the repelled
    species, for the repelled distribution of Fig.\ \ref{fvofvplot}.
    . \label{denplot}}
\end{figure}

The density is the integral $\int f_r(v_r) dv_r$ which is evidently
different for the two signs of $\sigma_z$, especially for small
$\varphi$, as illustrated for a typical $\varphi(z)$ shape in Fig.\
\ref{denplot}.
Notice that the attracted species density remains a function purely of
$\varphi$, and the deviation of the flat-trapped density $n_f-1$ (not
including the trapped depletion $\tilde n$) follows the shape of
$\varphi$.  The density difference results in an imbalanced net force
$F_r=\int n_r {-d\varphi\over dz} dz=\sum_{\sigma_z=\pm1}\int \sigma_zn_r
d\varphi$ exerted on the repelled species, whose reaction on the
potential itself must be balanced by momentum transfer to the
attracted species.

\subsection{Reflection force balance}

The dynamic behavior of a hole is determined by a balance of forces,
which can be written as $F_a+F_r=0$; that is: the sum of the forces
exerted by the hole on the attracted species plus the repelled species
must be zero. In essence, the momentum of the electrostatic field is
negligible; so it can support no reaction force. For electron holes of
small potential amplitude (${\psi}\ll T$), these two forces have been
calculated analytically in previous
publications\cite{Hutchinson2016,Hutchinson2017}. We here generalize
to ion holes and to non-equilibrium situations, though the analysis is
very similar. The mathematics is given only in outline, and relegated
to Appendix \ref{appb}.

It is helpful to distinguish contributions to the density difference
and force that arise from a possible potential difference
$\Delta \varphi=\varphi(+\infty)-\varphi(-\infty)$, which are called
extrinsic, from those, called intrinsic, which arise from asymmetries
in the distributions even in the absence of potential asymmetry. The
intrinsic force arises from the integral $\int(n_r-n_a) {d\varphi\over
  dz} dz$ over the $z$-range for which
$ \varphi\ge|\Delta\varphi|/2$. The extrinsic force comes from the
$z$-range over which
$-|\Delta\varphi|/2<\varphi<|\Delta\varphi|/2$, which is only on the low potential side
of the hole.

The attracted-species intrinsic force has been
shown\cite{Hutchinson2016} to be proportional to the acceleration
$\dot v_h$ (in units of $\sqrt{T_a/m_a}$) of the hole (in an inertial
frame) giving an effective Newton's second law that attributes the
momentum imparted to the trapped species to an effective mass:
\begin{equation}
  \label{Newtattr}
  \dot P_{aint} =M_a \dot v_h \qquad {\rm where} \qquad M_a=\int \tilde
  n dz.
\end{equation}
Here $\dot P_{aint}$ is the intrinsic rate of attracted species
momentum change, and $M_a$ is the effective mass of the hole. The mass
is negative, intuitively because the attracted species density deficit
$\tilde n$ is negative. The mass expression used here will be in a
form which correctly accounts for any reflected species influence on
the required density deficit ($\tilde n$), eq.\ (41) of
\citet{Hutchinson2016}, not the alternative form eq.\ (33) for a fast
electron hole in which ion perturbation can be ignored.

The repelled species intrinsic force arises to lowest order from
reflection. It is independent of acceleration and its net depends on
the antisymmetric part of the reflected velocity distribution function
in the hole frame of reference.  The resulting intrinsic momentum
transfer rate to (i.e.\ reflection force on) the repelled species is
\begin{equation}
  \label{Newtrepel}
  \dot P_{rint}=F_{rint}=\int_{\varphi\ge|\Delta\varphi|/2}
- n_r {d\varphi\over dz} dz,
\end{equation}
where we recall that $\varphi$ is normalized to $T_a/|q|$.

Approximating the distribution function, $f_{r\infty}$ (in repelled
species velocity units $\sqrt{T_a/m_r}$) by a Taylor expansion,
denoting differentiation with respect to velocity evaluated at the
hole velocity (e.g.\ $f_r'$, $f_r'''$), one can derive by integration
explicit expressions for $n_r$ and $F_{rint}$. See Appendix
\ref{appb}. Although the force expressions to be cited are inexact for
deep holes, ${\psi}\sim 1$, comparison with numerical calculation will
show that the approximation is very good.

The combined (repelled plus attracted) extrinsic force can be obtained
as the difference in the Maxwell stress across $\Delta\varphi$. The
electric field is zero at $\varphi=-|\Delta\varphi|$ (on the side with
lower $\varphi(\pm\infty)$), and, in view of the exponential potential
decay in this region, at $\varphi=+|\Delta\varphi|$ it is
$|E|=|\Delta\varphi|/\lambda$, giving a force (stress difference)
\begin{equation}
\dot P_{ext}=F_{ext}=-{\rm
  sign}(\Delta\varphi)|E|^2/2=-\Delta\varphi|\Delta\varphi|/2\lambda^2.  
\end{equation}
It is shown in the Appendix that, to first order in $\Delta\varphi$,
$F_{rint}+F_{ext}=\left[-4{\psi}^2 f_r'-{8\over9}{\psi}^3
  f_r'''\right]$.
So total momentum conservation is
\begin{equation}
0=\dot P_{aint}+F_{rint}+F_{ext}=M_a \dot v_h+\left[-4{\psi}^2
  f_r'-{8\over9}{\psi}^3 f_r'''\right].
\end{equation}

The potential difference across the hole $\Delta\varphi$ is required
to ensure distant neutrality, $n_r-n_a=0$ on both sides. Appendix \ref{appb}
shows it is
\begin{equation}\label{Delphi}
  \Delta\varphi= -\left(4{\psi}f'+{2\over3}\psi^2f'''\right)\Big/\left({1\over \lambda^2}-f'+{1\over12}{\psi}f'''\right).
\end{equation}

Equilibrium $\dot v_h=0$ requires $f_r'$ to be first order small
$f_r'= -{2\over9} f_r'''{\psi}$, and that occurs only at specific
discrete hole velocities $v_{eq}$ close to extrema in the distribution
shape. Here we concern ourselves in addition with non-equilibria in
which $\dot v_h\not=0$, and the acceleration will usually be well
approximated by an expression linear in the difference $v_h- v_{eq}$
between the hole velocity and a nearby equilibrium velocity.

\subsection{Numerical Calculation of Momentum Balance}

The alternative to the small-$\psi$ analysis of the previous
subsection is a full-scale numerical solution of the quasi-static
forces and hole mass. The code previously used to find asymmetric
electron hole equilibria and stability\cite{Hutchinson2021d}
accomplishes this task, but requires some modification to deal with
non-equilibrium situations or shifted attracted species Maxwellians.
In short, from a specified repelled distribution, in the hole frame,
it calculates the repelled density everywhere. It then determines from
the values $n_r(\pm\infty)$ the required $\Delta\varphi$ for distant
neutrality. The resulting forces on each species are $-\int nq
d\phi$. The $v_h$ can then be scanned, repeating this calculation, to
find the equilibrium velocity $v_{eq}$ at which the total force is
zero.

\begin{figure}
  \includegraphics[width=.49\hsize]{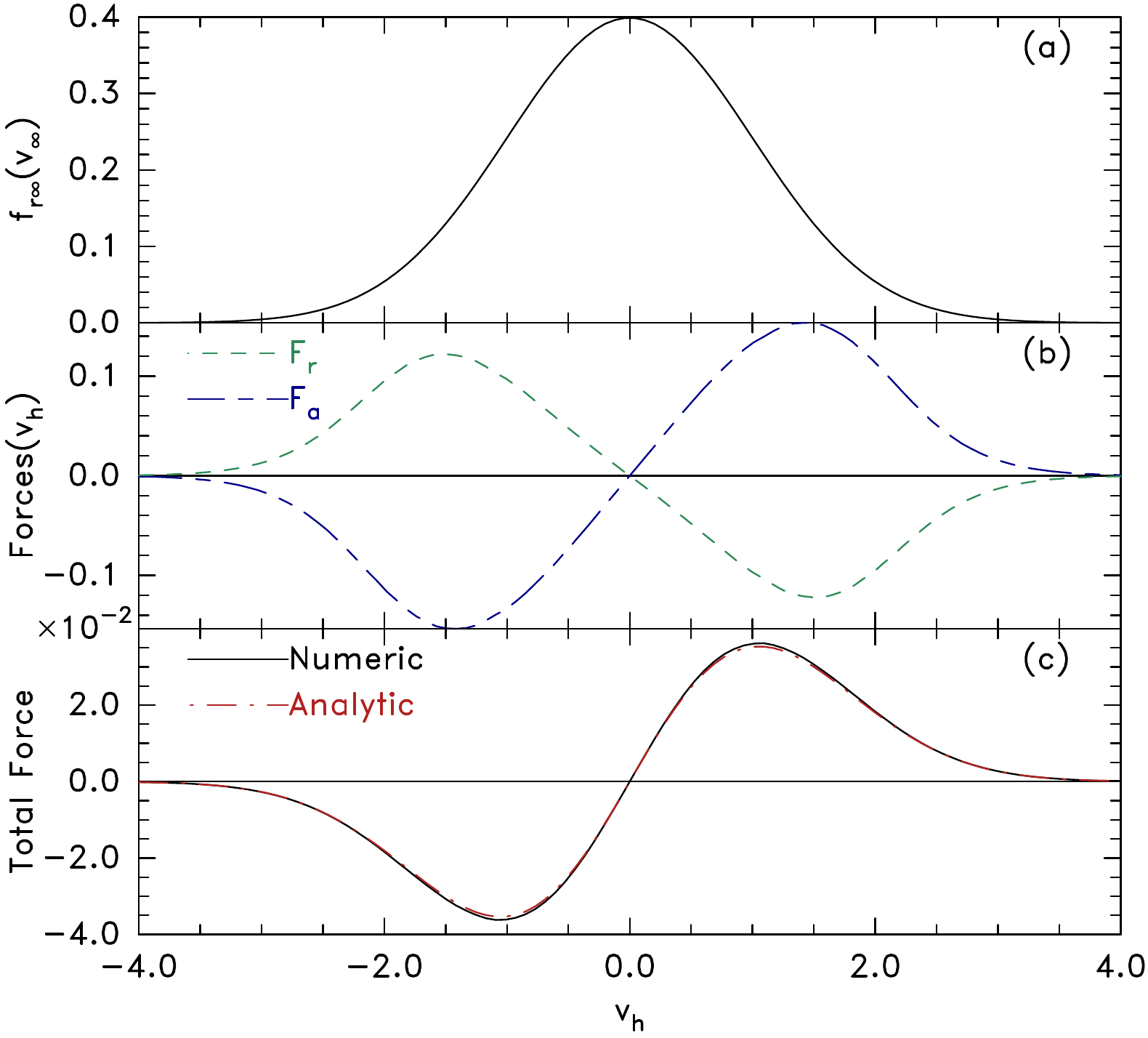}\hskip-2em(i)\hskip1em
  \includegraphics[width=.49\hsize]{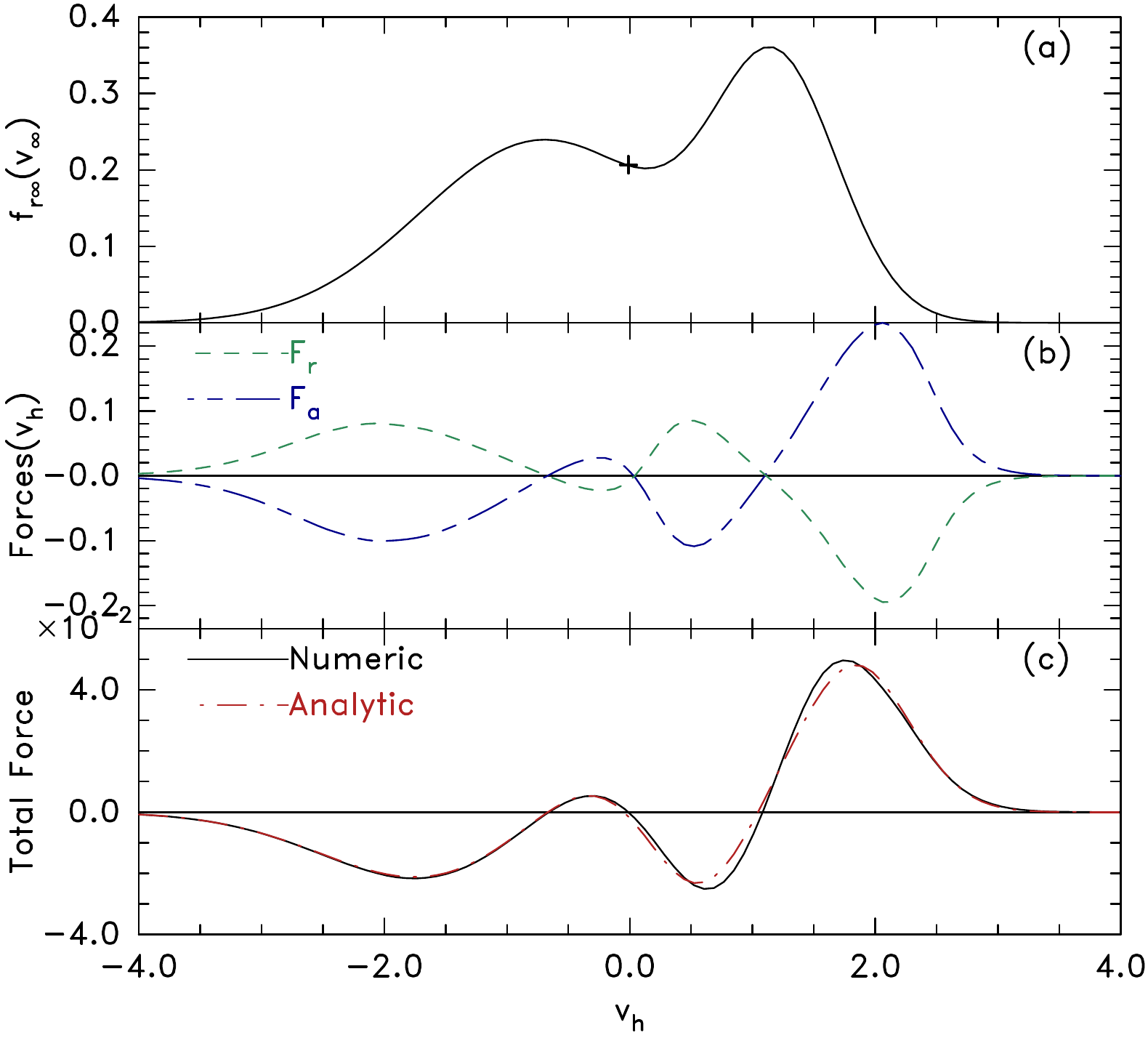}\hskip-2em(ii)\hskip1em
  \caption{Examples of force calculations showing: (a) the repelled
    distribution function vs.\ particle velocity, (b) vs.\ hole velocity
    the forces exerted on the repelled ($F_r$) and attracted ($F_a$)
    species, and (c) their sum, which is the total force, calculated
    numerically and compared with the analytic approximation
    $\left[-4{\psi}^2 f_r'-{8\over9}{\psi}^3 f_r'''\right]$.  The
    attracted species has a Maxwellian distribution unshifted relative
    to the hole velocity $v_h$. The hole amplitude is $\psi=0.2$.\label{fcomp}}
\end{figure}

Results are illustrated in Fig.\ \ref{fcomp}. In it velocities are
normalized to the repelled species, so for an ion hole the attracted
species distribution is far narrower and would appear to be a delta
function on the plot; in this calculation it is taken
as a Maxwellian centered on the hole velocity. 
At left, Fig.\ \ref{fcomp}(i), is shown the case for a simple
Maxwellian repelled distribution. The total force is zero when $v_h$
is at its centroid ($v_{eq}=0$), and rises initially linearly for
non-zero $v_h$ in a direction that would enhance $|v_h|$ by
acceleration. For large shifts the force returns to zero as the hole
disengages from the repelled species and reflection ceases.  At right,
Fig.\ \ref{fcomp}(ii), a double-humped repelled distribution is
illustrated. There are then three zeroes of the total force that would
correspond to equilibria. Those with positive force derivative are
unstable (as in Fig.\ \ref{fcomp}(i)), but the central zero with
negative derivative is stable. As for electron
holes\cite{Kamaletdinov2021,Hutchinson2021c}, stability generally
requires a local minimum in the repelled distribution and the hole
velocity lying within it.  The comparison between numerical and
analytic results shows how good the analytic approximation is at
$\psi=0.2$. For higher potential amplitudes it deteriorates, and for
lower amplitudes it becomes even better.

Nothing about the exact shape of the potential affects these
forces.
However, for non-equilibrium dynamics, the effective hole mass
$M_a=\int\tilde n dz$ depends upon the density spatial profile which
depends on the potential profile. So to find it we need the total
$\varphi(z)$. In the prior equilibrium publication the asymmetries
were always small and it was sufficient to use an approximation to the
$\varphi(z)$ on the unspecified side.  For strongly asymmetric holes
this is not as well justified. So, instead, to obtain the dependence
of $M_a$ on $ v_h$, the potential profile on the
higher-$\varphi(\infty)$ side ($\sigma_m$) is considered to be specified and
hence ${d^2\varphi\over dz^2}$. Thus
$\tilde n = n_r-n_f-{d^2\varphi\over dz^2}$ is found and eq.\
(\ref{eq:abel}) is evaluated to give $\tilde f$ on this side. At the
same $\varphi$ on the other side $\tilde f$ must be the same, because
the trapped orbits are the same. Therefore on the $-\sigma_m$ side
Poisson's equation can be solved with specified densities
$n(\varphi)$, to obtain $z(\varphi)$, complete the potential profile, and
evaluate $\int \tilde n dz$.

\begin{figure}\centering
  \includegraphics[width=.6\hsize]{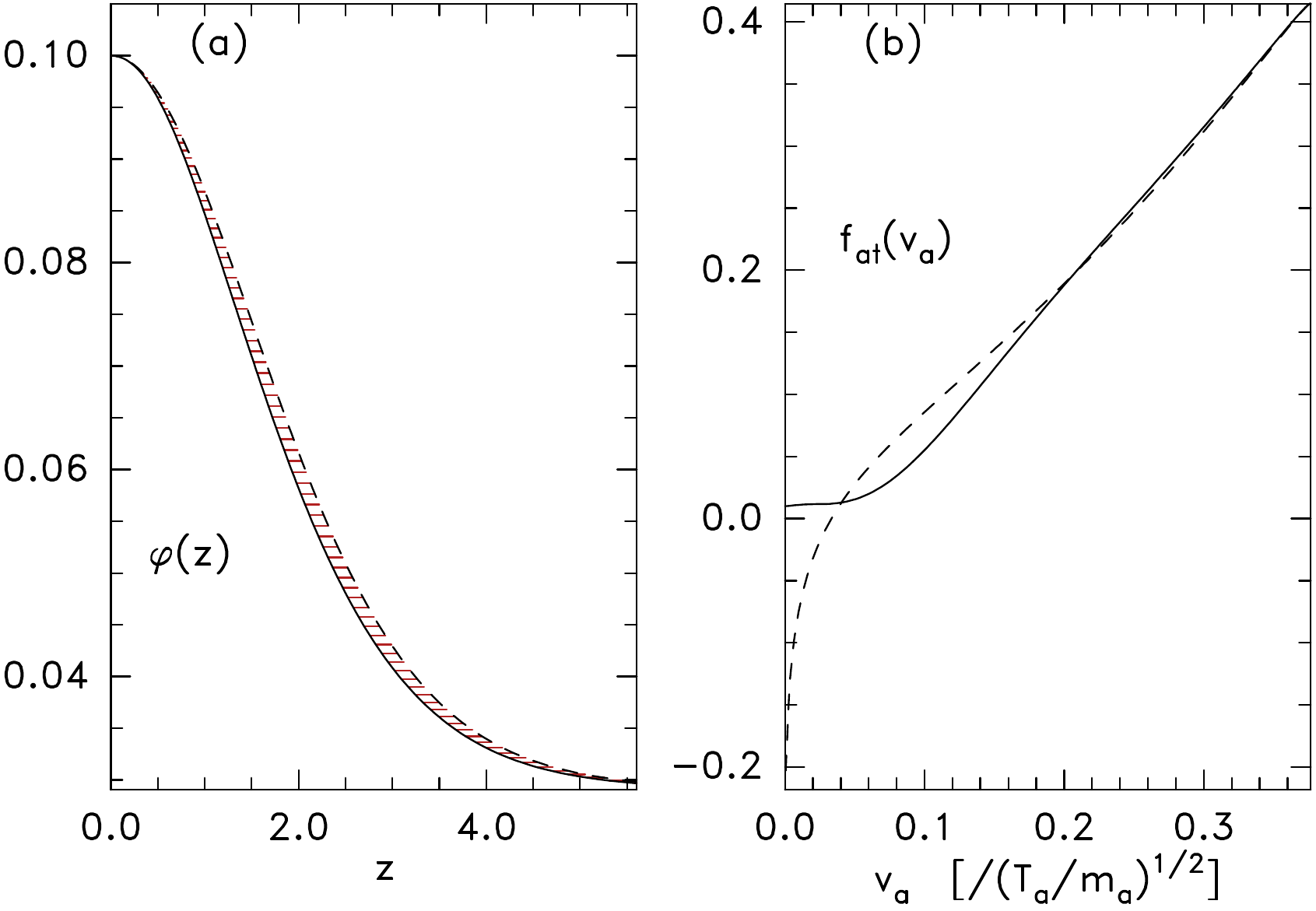}
  \caption{Potential profile $\varphi(z)$ and corresponding trapped
    distribution $f_{at}(v_a)$ on the higher distant potential side of
    a hole with a single Maxwellian repelled species shifted by
    $\bar v_r=-0.8$, when the attracted species is unshifted Maxwellian
    and $T_a/T_r=1$. Dashed lines are uncorrected and solid lines
    corrected by eq.\ (\ref{eq:cuspfix}). \label{denionslope}}
\end{figure}

A significant subtlety arises in the integral of eq.\ (\ref{eq:abel}) when
there is non-zero gradient ${dn_r\over dz}$ at the potential peak:
$z=0$, $\varphi=\psi$.  It is that since ${d\varphi\over dz}$ is zero
there, ${dn_r\over d\varphi}$ is singular. Both $n_r$ and $n_f$ are
bounded and have finite derivatives there; so if the potential shape
is such that ${d^3\varphi\over dz^3}={dn_\varphi\over dz}=0$, then
${d\tilde n\over dz}$ is singular, and the integration to give
$\tilde f$ results in an unphysical cusp at $z=0,\ v_a=0$. A simple
potential specification such as
$\varphi={\psi}{\rm sech}^\ell(z/\ell\lambda)$, for example, does give
${dn_\varphi\over dz}=0$ because, $\varphi$ being symmetric, has
${d^3\varphi\over dz^3}|_{z=0}=0$. It causes a cusp in $f_{at}(v_a)$, as
illustrated by the dashed lines in Fig.\ \ref{denionslope}. 
To avoid
this unphysicality, a potential shape must be chosen that gives rise
to a non-zero ${d^3\varphi\over dz^3}|_{z=0}$ value that exactly
cancels the ${dn_r\over dz}$. This requirement is, in effect, an
additional constraint on $\varphi$ profiles that are acceptable
physically. It can be satisfied, for example, by adopting a potential form
in which the $z$-variable is shifted by an increasing term:
\begin{equation}
  \label{eq:cuspfix}
  \varphi(z)={\psi}{\rm sech}^4(h(z))\qquad {\rm where}\qquad 
  h(z)={z\over 4\lambda} + {bz^2\over a^2+z^2},
\end{equation}
which has ${d^3\varphi\over dz^3}|_{z=0}=-{\psi 6 b\over \lambda a^2}$. We choose $a=1$ and
set the $b$ to cancel whatever ${dn_r\over dz}|_0$ is present. That
gives the solid lines in  Fig.\ \ref{denionslope}.

Actually this correction has little influence on the force and hole
mass calculation; and omitting it would be a minor blemish on our main
purpose. However, it appears to represent real physics, because it
amounts to applying approximately a shift of the hole's horizontal
center $\sum_{\sigma_z=\pm1}z(\varphi)/2$ by an amount that depends on
the height, being zero at the peak, but increasing to a constant at
large $|z|$. That shape change could be described intuitively as the
potential hill's peak ``leaning'' toward the side with higher
$n(\infty)$, but actually since the peak has been chosen to be at
$z=0$ the base of the potential profile shifts in the opposite
direction. These features are illustrated in Fig.\
\ref{denionslope}. Even for this extremely asymmetric hole
($v_h-\bar v_r=0.8$) the shape distortion in $\varphi$ is relatively
modest. It is the way the hole ``shares'' its effective inertia across
its entire extent, when it is being accelerated by asymmetric
reflection of the repelled species.  In the context of linearized
stability analysis, this effect causes a distortion of the
perturbation eigenstructure from the pure shiftmode.  For the
(two-dimensional) transverse instability, in which acceleration is
caused by shear stress rather than repelled species reflection, this
mode distortion has recently been treated via coupling of subsidiary
eigenmodes\cite{Chen2023}.

\subsection{Unstable acceleration}
When the hole velocity lies a little above an equilibrium value
$v_{eq}$ that is near a local $f_r$ maximum, the dominant term causing
acceleration, $-f_r'$, gives a positive force on the particles and the
reaction on the hole is negative. Then since $M_a$ is also negative,
the hole accelerates, moving its velocity away from the local $f_r$
peak.  Acceleration away from the peak also happens for a hole with
velocity just below a local $v_{eq}$. Consequently although a hole
with speed exactly at $v_{eq}$ associated with a $f_r$ maximum is in
equilibrium, any velocity perturbation $\Delta v_h=v_h-v_{eq}$ away
from $v_{eq}$ is unstable. On a single humped repelled velocity
distribution, an unstable electron hole can accelerate till repelled
species reflection ceases. But for an ion hole it disengages from the
attracted distribution, and consequently collapses in amplitude, long
before repelled species reflection ceases.

The growth rate of the instability near a local maximum at $v_{eq}$ is
determined to lowest order in $\psi$ by the second derivative of the
repelled distribution there $f_r''$, because
$f_r'(v_{eq}+\Delta v_h)\simeq f_r'(v_{eq})+f_r''(v_{eq}) \Delta v_h$. Therefore to
lowest order in $\Delta v_h$
\begin{equation}
  \label{growth}
  \dot v_h={d\Delta v_h\over dt}  \simeq 4{\psi}^2f_r''(v_{eq})\Delta v_h/M_a,
\end{equation}
giving an exponential growth $\Delta v_h\propto {\rm e}^{\gamma t}$,
neglecting the dependence of $M_a$ on $v_h$, with growth rate
$\gamma=4{\psi}^2f_r''(v_{eq})/M_a$. Of course,
near a local $f_r$ minimum where $f_r''$ is positive, by contrast,
$\gamma$ is negative, and in this approximation the hole velocity
shift decays with time towards zero where $v_h=v_{eq}$. 

Notice that this treatment regards the repelled distribution as
exerting a force on an accelerating hole that is independent of the
acceleration. That is hardly justifiable when the repelled species is
ions because the timescale for repelled force ion equilibration is
much longer than the typical $1/\gamma$. That is one reason why slow
electron holes need a more elaborate treatment\cite{Hutchinson2022}.
Ion holes, by contrast, accelerate far slower than electrons
equilibrate; so the repelled species static approximation is very well
justified.

In absolute units the $f_r''$ being far smaller for ion holes than
electron holes reduces the growth rate enough to justify the attracted
species approximations as follows. In eq.\ (\ref{growth}) the time
unit in $d\Delta v_h/dt$ is $\omega_{pa}^{-1}$, but the velocity unit
in $f_r$ and its derivatives is $\sqrt{T_a/m_r}$ which is large when
the repelled species is electrons. Therefore a Maxwellian repelled
distribution with temperature $T_r$ has, at its peak,
$f_r''=-(T_a/T_r)^{3/2}/\sqrt{2\pi}=-\theta^{-3/2}/\sqrt{2\pi}$, and
the instability growth rate is
$\gamma= -(4/\sqrt{2\pi}){\psi}^2\theta^{-3/2}/M_a$
[$\times \omega_{pa}$ in absolute units]. Since the hole mass ($M_a$)
units are the same for either species, the growth rate scaling like
${\psi}^2\theta^{-3/2}/M_a$ attracted species time units gives growth
rate slower in absolute time units for ion holes than electron holes
by a factor $\sim\sqrt{m_e/m_i}$. Consequently the approximations for
the attracted species are approximately as well justified for ion
holes as they are for electron holes.

Quantitatively, taking $\ell=4$, so $\varphi=\psi\sech^4(z/4\lambda)$ which is a
representative choice, and unshifted Maxwellians, a small-${\psi}$
expansion gives
\begin{equation}
  \label{intn}
  M_a=\int \tilde n dz = \left[-{16\over 3}(1+1/\theta){\psi}
  -{512\over 45\sqrt{\pi}}{\psi}^{3/2}
  +{64\over 35\theta^2}{\psi}^2\right](1+1/\theta)^{-1/2}+O({\psi}^{5/2}),
\end{equation}
which to leading order is
$M_a=-(16/3){\psi}\sqrt{1+1/\theta}$,
giving
\begin{equation}
  \label{gammafinal}
  \gamma\simeq{3\over4\sqrt{2\pi}}{{\psi}\over \theta\sqrt{\theta+1}}
.
\end{equation}
For a small initial perturbation from equilibrium $\Delta v_{h0}$
to grow to of order unity, taking the hole to the edge of the
attracted velocity distribution, then requires a time $t\simeq-\ln (\Delta
v_{h0})/\gamma$ which for 10 e-foldings ($\Delta v_{h0}\sim {\rm
  e}^{-10}$) is $t_{loss}\sim 10/\gamma\simeq 33
\theta\sqrt{\theta +1}/{\psi}$ ($\omega_{pa}^{-1}$). 

Since the attracted deeply trapped particle bounce frequency is
$\omega_b \sim \sqrt{\psi}/2$ ($\omega_{pa}$), there are many bounces in the time it
takes for the hole to be accelerated to a velocity that causes it to
dissipate:
$\omega_b t_{loss}\sim33 \theta\sqrt{\theta +1}/\sqrt{{\psi}} \gg
1$. This inequality indicates that even though an ion  hole is unstable in
a Maxwellian background, it still retains its integrity long enough that
the trapped dynamics is well represented by the equations we have been
using.  Its lifetime is also long enough to be observed as virtually
steady (in its moving frame) by a satellite transiting at speeds
at least comparable to the trapped particles.

Naturally if an ion hole lies in a locally flat $f_r$ region or local
minimum, its velocity is stable by the present analysis, and its
lifetime could be much longer. We have not here addressed
whether there are oscillating instabilities\cite{Hutchinson2022}, even
when the hole lies in a minimum of $f_r$, arising from small
phase-shifts in the attracted species response.

If an unstable ion hole is somehow born (created) at already
substantial $\Delta v_h\sim 1$ then only approximately one e-folding
would be sufficient for it to dissipate by accelerating beyond the
attracted distribution extent, in a time
$t_{loss} \sim 3 \theta\sqrt{\theta +1}/\sqrt{{\psi}}$. This might
well be the relevant limit on hole life.  However, provided the
temperature ratio $\theta=T_r/T_a$ is not too small, it would still be
long enough for a typical satellite transit (at well above the ion
thermal speed) to observe an approximately steady ion hole. Therefore
the mere observation of an ion hole by a satellite does not rule out
its accelerating at the rates calculated here.

Moreover, the growth rate being inversely proportional to
$\theta\sqrt{\theta+1}$ means it is reduced for an ion hole when the
electron temperature is increased. Thus ion holes in higher $T_e$
plasmas last longer before being collapsed by electron reflection
acceleration. This effect is consistent with the tendency for ion
holes to be more frequently observed in plasmas with higher
$T_i/T_e$\cite{Wang2021}, which is sometimes supposed to result from
the erroneous $\theta$ limit discussed in section \ref{erroneous}.

\section{Discussion}\label{sec4}

The phenomenon of transverse instability, which is known to be vital
for electron holes, has not been addressed here for ion holes. There
is every theoretical reason to suppose that ion holes are even more
subject to it, because the effects of magnetization on the ion orbits
are much reduced in the same plasma. That is because the ratio of ion
plasma frequency to gyro frequency $\omega_{pi}/\Omega_i$ is greater
by the square root of the mass ratio than $\omega_{pe}/\Omega_e$. Thus
in space plasmas ion holes hardly ever reside in a magnetic field
strong enough that that $\Omega_i$ exceeds approximately 1.5 times the
ion bounce frequency $\omega_b\sim \omega_{pi}\sqrt{\psi}/2$. Instead
the ions are effectively unmagnetized. The unmagnetized transverse
instability fastest growth rate
is\cite{Hutchinson2018,Hutchinson2018a} approximately $\sqrt{\psi}/16$
($\omega_{pi}$ units for an ion hole). This exceeds the
self-acceleration growth rate
$\gamma\simeq 0.3\psi/(\theta\sqrt{\theta+1})$ when
$\psi\lesssim 0.2\,\theta^2(\theta+1)$. Thus small amplitude ion holes
appear more likely to be limited by the transverse instability than
self-acceleration, if they start from an equilibrium. An ion hole born
already at a speed with substantial electron distribution gradient
($f'$) would, however, accelerate promptly without having to wait for
instability growth from noise, and if its amplitude is as great as
$\psi\sim 0.1$ might well collapse by acceleration before transverse
instability breakup.

Although we have illustrated non-Maxwellian calculations, most of the
explicit formulas here developed are given for Maxwellian
distributions. That is not so great a weakness as might be supposed
for space plasmas, even though they often have non-Maxwellian tails to
their distributions, sometimes modelled by kappa functions (e.g.\
\cite{Haas2021}). The reason is that electron influence on ion holes
arises primarily from reflection, which occurs only for electron
velocities close to the hole velocity, lying in the bulk of the
electron distribution. Energetic tails on the wings of the electron
distribution only marginally affect the hole equilibrium and give zero
reflection force tending to accelerate it. Tails on the ion
distribution might permit ion holes to reach higher speed before
dissipating because of the small background distribution in which they
are holes, but they will still quite soon be limited by the mechanisms
giving rise to Fig.\ \ref{fig:ftildef}(b), after some modest extension
in $v_a$.

It should be emphasized that simulating ion holes by treating
electrons via a fluid approximation, for example a Boltzmann density
approximation, generally omits the asymmetric repelled species
reflection effects that are the cause of ion hole acceleration
discussed here. Thus, the electrons must include kinetic effects in
their representation. The present analysis indicates that ion hole dynamics
are affected only relatively little by the alternative simulation
strategy of using a kinetic treatment of both species with unphysical
increased electron mass to mitigate computational demands (e.g.\
\cite{Berman1985,Lesur2014}). For example, the unstable hole velocity
growth rate (eq.\ \ref{gammafinal}) is unaffected, and ion holes rarely
disengage with electrons. The same cannot be said about electron hole
dynamics. They are greatly influenced by the mass ratio, since it
dictates the velocity at which electron holes disengage from ions and
cease to be accelerated.

The presumption made here that asymmetric background electron
distribution functions are prescribed, becomes questionable for trains
of ion holes, sometimes described as Cnoidal
waves\cite{Schamel2018}. Electrons that reflect from one hole also
reflect from an adjacent hole, and are therefore themselves trapped
between two $\varphi$ peaks. Their phase-space density is therefore
not prescribed by some distant boundary condition. Trapping causes the
reflecting electron orbits' distribution to be symmetrized  in
velocity. And (in contast with electron holes) this repelled species
symmetrization takes place on the electron transit timescale, which is
much shorter than the ion timescale. In a train of similar holes
making up a periodic wave, the result suppresses any reflection
asymmetry force on an individual ion hole, and thus its tendency to
accelerate. Addressing this situation is beyond the present scope.

\textbf{To summarize,} the equations governing equilibrium and
dynamics of ion holes have been derived and solved analytically and
numerically for representative potential shapes and
distributions. Most of the mathematics is the same as for electron
holes when expressed in appropriately normalized units. Ion holes can
exist for a wide range of temperature ratios $\theta=T_r/T_a$, in
contradiction of an erroneous but widely cited criterion. However, ion
holes almost always give rise to electron reflection, in which case
they are accelerated by asymmetry in the electron velocity
distribution. Their lifetime is thereby limited to at most the time
taken to accelerate out of the background ion velocity distribution,
of order $33\theta\sqrt{\theta+1}/\sqrt{\psi}\
(\omega_{pi}^{-1})$. This is nevertheless long enough for the
quasi-static analyis presented to be reasonably appropriate, and for
satellites to observe ion holes as approximately static in their rest
frame.

\appendix
\section{Waterbag Deficit Model}\label{appa}

The ``power deficit'' model\cite{Hutchinson2021a} consists of the assumption that the phase
space density deficit is a specified function of energy
\begin{equation}
  \tilde f(\energy) \propto (\energy_j-\energy)^\alpha
\end{equation}
for $\energy < \energy_j$ and zero for orbit energy
$\energy \ge \energy_j$, where $\energy_j$ is a non-positive constant
and $\alpha\ge 0$ is the chosen power.  This form allows for variable
shape, but the enforced flat trapped $f$ near the trapping boundary
accounts for possible rapid energy diffusion there caused by
stochasticity\cite{Hutchinson2020}. One can regard it as a
generalization of the form used by \citet{Bohm1949}, which had
$\alpha=1/2$, $\energy_j=0$. Notation is simplified by adopting the
convention that any negative quantity to a real power (such as
$(\energy_j-\energy)^\alpha$ when $\energy > \energy_j$) is zero. The
mathematical convenience of this power form is that it can be
integrated to give
\begin{equation}
\tilde n(\varphi)=2 G {\tilde f(-\psi)\over
(\psi+\energy_j)^\alpha  }(\varphi+\energy_j)^{\alpha+1/2},
\end{equation}
where $G=\sqrt{\pi/2}\,\Gamma(\alpha+1)/\Gamma(\alpha+3/2)$, and
$\Gamma$ is the standard Gamma function. This expression can further
be integrated as 
\begin{equation}
  \int \tilde n d\varphi=2 G {\tilde f(-\psi)\over
(\psi+\energy_j)^\alpha  }{(\varphi+\energy_j)^{\alpha+3/2}\over
\alpha +3/2}
\end{equation}
For small amplitude holes, we may approximate the rest of the density
as $n_f(\varphi)-n_r(\varphi)=K\varphi$, where $K$ is a positive constant. 
Then a closed form first integral of Poisson's equation is
\begin{equation}
  \label{firstint}
  -\left({d\varphi\over dz}\right)^2=2\int (n_r-n_f-\tilde n)d\varphi
=-K\varphi^2-2 G {\tilde f(-\psi)\over
(\psi+\energy_j)^\alpha  }{(\varphi+\energy_j)^{\alpha+3/2}\over
\alpha +3/2}.
\end{equation}
The constant of integration is zero because $\left.d\varphi\over dz\right|_{\varphi=0}=0$.
At the potential peak ($\varphi=\psi$), ${d\varphi\over dz}$ is again zero so 
\begin{equation}
  \label{tildefpsi}
  \tilde f(-\psi)/\psi^{1/2} = -K{\alpha+3/2\over4G} (1+\energy_j/\psi)^{-3/2}.
\end{equation}
But the peak will be reached only if the derivative of the right
hand side of eq.\ (\ref{firstint}) remains positive. Substituting from 
eq.\ (\ref{tildefpsi}) into that derivative we obtain the condition
$K\psi(-2+(\alpha+3/2)(1+\energy_j/\psi)>0 $,
which is 
\begin{equation}
  \energy_j/\psi < (\alpha-1/2)/2.
\end{equation}
This shows that for $\alpha\to 0$, which corresponds to a step
function with transition at $\energy_j$, i.e. a ``waterbag'' shape, the
maximum permitted value of $\energy_j/\psi$ is $-1/4$. This case,
$\alpha=0$, $\energy_j/\psi=-1/4$, permits the highest value of $\psi$
subject to the non-negativity constraint
$\tilde f(-\psi) \ge -f_a(0)$. And the value is then
\begin{equation}
  \psi_{max}= 3(Gf_a(0)/K)^2.
\end{equation}
Substituting $G=\sqrt{2}$ for $\alpha=0$, $f_a(0)=1/\sqrt{2\pi}$ for an
unshifted Maxwellian, and $K=1+T_a/T_r$ to account for the response of
both species, we obtain 
\begin{equation}
  \psi_{max}={3\over\pi(1+T_a/T_r)^2}.
\end{equation}

\section{Analytic Density Difference and Force Expressions}\label{appb}

Insofar as the antisymmetric part of the repelled species density can
be represented through the first two nonzero terms in its
distribution's Taylor series in velocity:
$\Delta f_r= -(2f' v_\energy+{1\over 3}f'''v_\energy^3)$ --- where for
any potential or energy quantity (such as $\energy$) we use the
notation $v_\energy\equiv \sqrt{2\energy}$ --- it is straightforward to
show by integration [see \citet{Hutchinson2021d} eq.\ 15] that the
corresponding difference in repelled density between the two sides of
the hole, at potential $\varphi$ such that
$|\Delta\varphi|/2\le \varphi\le {\psi}$, is
\begin{equation}
  \label{eq:Dnint}
  \begin{split}
  \Delta n_{int}(\varphi) =& -2\left[v_{{\psi}-\varphi} v_{{\psi}} +v_\varphi^2\ln({v_{{\psi}-\varphi}+v_{{\psi}}\over v_\varphi})\right]f'\\
  &-{1\over 12}\left[v_{{\psi}-\varphi} v_{{\psi}}(2v_{{\psi}}^2+3v_\varphi^2)+3v_\varphi^4 \ln({v_{{\psi}-\varphi}+v_{{\psi}}\over v_\varphi})\right]f'''.
  \end{split}
\end{equation}
The corresponding intrinsic force (difference) contributed from
potentials above
$\varphi$ is
$F_{rint}(\varphi)=\int_\varphi^\psi \Delta n_{int}(\varphi) d\varphi = {1\over
  2}\int_{v_{\varphi}^2}^{v_\psi^2} \Delta n_{int}\; dv_\varphi^2$
and the integrals can be evaluated to give
\begin{equation}
  \label{eq:int2}\begin{split}
    F_{rint}(\varphi)  =&-{1\over 2}\left[
      v_{{{\psi}}-\varphi}v_{{\psi}}(-2v_{{\psi}}^2+v_\varphi^2)+
      v_\varphi^4\ln({v_{{{\psi}}-\varphi}+v_{{\psi}}\over v_\varphi})\right]f'
    \\
    &-{1\over 24}\left[{1\over3}
      v_{{{\psi}}-\varphi}v_{{\psi}}(-8v_{{\psi}}^4+2v_{{\psi}}^2
      v_\varphi^2+3v_\varphi^4)
      +v_\varphi^6\ln({v_{{{\psi}}-\varphi}+v_{{\psi}}\over
        v_\varphi})\right]f'''.
  \end{split}
\end{equation}
At $\varphi=0$ this expression gives
$F_{rint}(0)=-v_{{\psi}}^4f' -{1\over 9}v_{{\psi}}^6f'''$; and in
equilibrium situations when $\Delta\varphi$ is small, this is a
sufficient approximation.  However, since the higher potential side of
the hill descends only to
$\varphi=\varphi_\Delta\equiv |\Delta\varphi|/2$, the intrinsic force
is actually $F_{rint}(\varphi_\Delta)$ and far from equilibrium a
more accurate expression is needed. The difference between the two
expressions to first order in $\Delta\varphi$ is
\begin{equation}\label{frintdiff}
  F_{rint}(0)-F_{rint}(\varphi_\Delta)=-2\psi|\Delta\varphi|f'
  -{1\over3}\psi^2|\Delta\varphi|f'''.
\end{equation}
And for non-equilibrium, accelerating, conditions this
correction is required.

The total extrinsic force is
$F_{ext}=\int_{-\Delta\varphi/2}^{\Delta\varphi/2} (n_r-n_a)d\varphi$,
and if we approximate the densities as decaying with constant
$\varphi$-derivative, giving $n_r-n_a=0$ at
$\varphi=-\varphi_\Delta$
 then
 \begin{equation}
n_r(\varphi)-n_a(\varphi)={n_r(\varphi_\Delta)-n_a(\varphi_\Delta)\over\Delta\varphi}(\varphi+\varphi_\Delta),
 \end{equation}
and the integral
results in
\begin{equation}
F_{ext}=\Delta n(\varphi_\Delta)
\varphi_\Delta=-2|\psi\Delta\varphi| f'-{1\over 3}|\psi^2\Delta\varphi|
f'''.  
\end{equation}
This expression is precisely equal to the difference expression eq.\
(\ref{frintdiff}). Consequently, to first order in $\Delta\varphi$,
the difference is cancelled and total force is
\begin{equation}
  \label{totalforce}
  F_{rint}(|\Delta\varphi|/2)+F_{ext}=F_{rint}(0)=-4\psi^2f'-{1\over9}{\psi}^3f''',
\end{equation}
somewhat surprisingly independent of $\Delta\varphi$.

The intrinsic repelled density difference between the two sides of the
hole at $\varphi_\Delta$ where the sign of $z$ is $\sigma_z$ is
obtained to first order in $\Delta\varphi$ from eq.\ (\ref{eq:Dnint})
as
\begin{equation}
\sum_{\sigma_z=\pm1}  \sigma_zn_r(\varphi_\Delta)=  \Delta n_{int}(\varphi_\Delta)=-(4{\psi}+|\Delta\varphi|)f'
-{\psi}\left({2\over3}{\psi}+{1\over12}|\Delta\varphi|\right)f'''.
\end{equation}
Ignoring acceleration, there is no intrinsic difference in the
attracted species density between the two sides of the hole at
$\varphi_\Delta$. The ``external'' difference between repelled and
attracted densities at this position can also be approximated to first
order using eq.\ (\ref{eq:shielding}) and external neutrality at the
lower side $n_r(-\varphi_\Delta)-n_a(-\varphi_\Delta)=0$ as
\begin{equation}
\Delta n_{ext}(\varphi_\Delta)= {\rm sign}(\Delta\varphi)[ n_r(\varphi_\Delta)-n_a(\varphi_\Delta)]\simeq
({dn_r\over d\varphi}-{dn_a\over d\varphi})\Delta\varphi=
-{\Delta\varphi\over \lambda^2}.
\end{equation}
Then neutrality at the higher $\varphi(\infty)$ side requires $\Delta
n_{int}+\Delta n_{ext}=0$, which is
\begin{equation}
  -(4{\psi}f'+{2\over3}\psi^2f''')+\Delta\varphi(f'-{1\over12}{\psi}f'''-{1\over \lambda^2})=0,
\end{equation}
giving eq.\ (\ref{Delphi}).

\section*{Acknowledgments}

I am grateful to Ivan Vasko for helpful conversations. The code used
to calculate and plot the figures is openly available at
 \url{https://github.com/ihutch/asymhill}.

\bibliography{JabRef}

\begin{thebibliography}{46}%
\makeatletter
\providecommand \@ifxundefined [1]{%
 \@ifx{#1\undefined}
}%
\providecommand \@ifnum [1]{%
 \ifnum #1\expandafter \@firstoftwo
 \else \expandafter \@secondoftwo
 \fi
}%
\providecommand \@ifx [1]{%
 \ifx #1\expandafter \@firstoftwo
 \else \expandafter \@secondoftwo
 \fi
}%
\providecommand \natexlab [1]{#1}%
\providecommand \enquote  [1]{``#1''}%
\providecommand \bibnamefont  [1]{#1}%
\providecommand \bibfnamefont [1]{#1}%
\providecommand \citenamefont [1]{#1}%
\providecommand \href@noop [0]{\@secondoftwo}%
\providecommand \href [0]{\begingroup \@sanitize@url \@href}%
\providecommand \@href[1]{\@@startlink{#1}\@@href}%
\providecommand \@@href[1]{\endgroup#1\@@endlink}%
\providecommand \@sanitize@url [0]{\catcode `\\12\catcode `\$12\catcode
  `\&12\catcode `\#12\catcode `\^12\catcode `\_12\catcode `\%12\relax}%
\providecommand \@@startlink[1]{}%
\providecommand \@@endlink[0]{}%
\providecommand \url  [0]{\begingroup\@sanitize@url \@url }%
\providecommand \@url [1]{\endgroup\@href {#1}{\urlprefix }}%
\providecommand \urlprefix  [0]{URL }%
\providecommand \Eprint [0]{\href }%
\providecommand \doibase [0]{https://doi.org/}%
\providecommand \selectlanguage [0]{\@gobble}%
\providecommand \bibinfo  [0]{\@secondoftwo}%
\providecommand \bibfield  [0]{\@secondoftwo}%
\providecommand \translation [1]{[#1]}%
\providecommand \BibitemOpen [0]{}%
\providecommand \bibitemStop [0]{}%
\providecommand \bibitemNoStop [0]{.\EOS\space}%
\providecommand \EOS [0]{\spacefactor3000\relax}%
\providecommand \BibitemShut  [1]{\csname bibitem#1\endcsname}%
\let\auto@bib@innerbib\@empty
\bibitem [{\citenamefont {Bernstein}, \citenamefont {Greene},\ and\
  \citenamefont {Kruskal}(1957)}]{Bernstein1957}%
  \BibitemOpen
  \bibfield  {author} {\bibinfo {author} {\bibfnamefont {I.~B.}\ \bibnamefont
  {Bernstein}}, \bibinfo {author} {\bibfnamefont {J.~M.}\ \bibnamefont
  {Greene}},\ and\ \bibinfo {author} {\bibfnamefont {M.~D.}\ \bibnamefont
  {Kruskal}},\ }\bibfield  {title} {\enquote {\bibinfo {title} {{Exact
  nonlinear plasma oscillations}},}\ }\href
  {http://journals.aps.org/pr/abstract/10.1103/PhysRev.108.546} {\bibfield
  {journal} {\bibinfo  {journal} {Physical Review}\ }\textbf {\bibinfo {volume}
  {108}},\ \bibinfo {pages} {546--550} (\bibinfo {year} {1957})}\BibitemShut
  {NoStop}%
\bibitem [{\citenamefont {Bujarbarua}\ and\ \citenamefont
  {Schamel}(1981)}]{Bujarbarua1981}%
  \BibitemOpen
  \bibfield  {author} {\bibinfo {author} {\bibfnamefont {S.}~\bibnamefont
  {Bujarbarua}}\ and\ \bibinfo {author} {\bibfnamefont {H.}~\bibnamefont
  {Schamel}},\ }\bibfield  {title} {\enquote {\bibinfo {title} {Theory of
  finite-amplitude electron and ion holes},}\ }\href@noop {} {\bibfield
  {journal} {\bibinfo  {journal} {Journal of Plasma Physics}\ }\textbf
  {\bibinfo {volume} {25}},\ \bibinfo {pages} {515--529} (\bibinfo {year}
  {1981})}\BibitemShut {NoStop}%
\bibitem [{\citenamefont {Schamel}(1982)}]{Schamel1982}%
  \BibitemOpen
  \bibfield  {author} {\bibinfo {author} {\bibfnamefont {H.}~\bibnamefont
  {Schamel}},\ }\bibfield  {title} {\enquote {\bibinfo {title} {{Stability of
  electron vortex structures in phase space}},}\ }\href@noop {} {\bibfield
  {journal} {\bibinfo  {journal} {Physical Review Letters}\ }\textbf {\bibinfo
  {volume} {48}},\ \bibinfo {pages} {481--483} (\bibinfo {year}
  {1982})}\BibitemShut {NoStop}%
\bibitem [{\citenamefont {Dupree}(1982)}]{Dupree1982}%
  \BibitemOpen
  \bibfield  {author} {\bibinfo {author} {\bibfnamefont {T.~H.}\ \bibnamefont
  {Dupree}},\ }\bibfield  {title} {\enquote {\bibinfo {title} {{Theory of
  phase-space density holes}},}\ }\href {https://doi.org/10.1063/1.863734}
  {\bibfield  {journal} {\bibinfo  {journal} {Physics of Fluids}\ }\textbf
  {\bibinfo {volume} {25}},\ \bibinfo {pages} {277} (\bibinfo {year}
  {1982})}\BibitemShut {NoStop}%
\bibitem [{\citenamefont {Pecseli}, \citenamefont {Armstrong},\ and\
  \citenamefont {Trulsen}(1981)}]{Pecseli1981}%
  \BibitemOpen
  \bibfield  {author} {\bibinfo {author} {\bibfnamefont {H.}~\bibnamefont
  {Pecseli}}, \bibinfo {author} {\bibfnamefont {R.}~\bibnamefont {Armstrong}},\
  and\ \bibinfo {author} {\bibfnamefont {J.}~\bibnamefont {Trulsen}},\
  }\bibfield  {title} {\enquote {\bibinfo {title} {Experimental observations of
  ion phase-space vortices},}\ }\href@noop {} {\bibfield  {journal} {\bibinfo
  {journal} {Physics Letters A}\ }\textbf {\bibinfo {volume} {81}},\ \bibinfo
  {pages} {386--390} (\bibinfo {year} {1981})}\BibitemShut {NoStop}%
\bibitem [{\citenamefont {Johnsen}, \citenamefont {Pécseli},\ and\
  \citenamefont {Trulsen}(1987)}]{Johnsen1987}%
  \BibitemOpen
  \bibfield  {author} {\bibinfo {author} {\bibfnamefont {H.}~\bibnamefont
  {Johnsen}}, \bibinfo {author} {\bibfnamefont {H.~L.}\ \bibnamefont
  {Pécseli}},\ and\ \bibinfo {author} {\bibfnamefont {J.}~\bibnamefont
  {Trulsen}},\ }\bibfield  {title} {\enquote {\bibinfo {title} {Conditional
  eddies in plasma turbulence},}\ }\href
  {https://doi.org/.org/10.1063/1.2355668} {\bibfield  {journal} {\bibinfo
  {journal} {Physics of Fluids}\ }\textbf {\bibinfo {volume} {30}},\ \bibinfo
  {pages} {2239--2254} (\bibinfo {year} {1987})}\BibitemShut {NoStop}%
\bibitem [{\citenamefont {Sakanaka}(1972)}]{Sakanaka1972}%
  \BibitemOpen
  \bibfield  {author} {\bibinfo {author} {\bibfnamefont {P.~H.}\ \bibnamefont
  {Sakanaka}},\ }\bibfield  {title} {\enquote {\bibinfo {title}
  {Beam‐generated collisionless ion‐acoustic shocks},}\ }\href
  {https://doi.org/.org/10.1063/1.859653} {\bibfield  {journal} {\bibinfo
  {journal} {Physics of Fluids}\ }\textbf {\bibinfo {volume} {15}},\ \bibinfo
  {pages} {1323--1327} (\bibinfo {year} {1972})}\BibitemShut {NoStop}%
\bibitem [{\citenamefont {Pecseli}, \citenamefont {Trulsen},\ and\
  \citenamefont {Armstrong}(1984)}]{Pecseli1984}%
  \BibitemOpen
  \bibfield  {author} {\bibinfo {author} {\bibfnamefont {H.}~\bibnamefont
  {Pecseli}}, \bibinfo {author} {\bibfnamefont {J.}~\bibnamefont {Trulsen}},\
  and\ \bibinfo {author} {\bibfnamefont {R.}~\bibnamefont {Armstrong}},\
  }\bibfield  {title} {\enquote {\bibinfo {title} {Formation of ion phase-space
  vortexes},}\ }\href@noop {} {\bibfield  {journal} {\bibinfo  {journal}
  {Physica Scripta}\ }\textbf {\bibinfo {volume} {29}},\ \bibinfo {pages}
  {241--253} (\bibinfo {year} {1984})}\BibitemShut {NoStop}%
\bibitem [{\citenamefont {Berman}, \citenamefont {Tetreault},\ and\
  \citenamefont {Dupree}(1985)}]{Berman1985}%
  \BibitemOpen
  \bibfield  {author} {\bibinfo {author} {\bibfnamefont {R.~H.}\ \bibnamefont
  {Berman}}, \bibinfo {author} {\bibfnamefont {D.~J.}\ \bibnamefont
  {Tetreault}},\ and\ \bibinfo {author} {\bibfnamefont {T.~H.}\ \bibnamefont
  {Dupree}},\ }\bibfield  {title} {\enquote {\bibinfo {title} {{Simulation of
  phase space hole growth and the development of intermittent plasma
  turbulence}},}\ }\href {https://doi.org/10.1063/1.865176} {\bibfield
  {journal} {\bibinfo  {journal} {Physics of Fluids}\ }\textbf {\bibinfo
  {volume} {28}},\ \bibinfo {pages} {155--176} (\bibinfo {year}
  {1985})}\BibitemShut {NoStop}%
\bibitem [{\citenamefont {Goldman}, \citenamefont {Newman},\ and\ \citenamefont
  {Ergun}(2003)}]{Goldman2003}%
  \BibitemOpen
  \bibfield  {author} {\bibinfo {author} {\bibfnamefont {M.~V.}\ \bibnamefont
  {Goldman}}, \bibinfo {author} {\bibfnamefont {D.~L.}\ \bibnamefont
  {Newman}},\ and\ \bibinfo {author} {\bibfnamefont {R.~E.}\ \bibnamefont
  {Ergun}},\ }\bibfield  {title} {\enquote {\bibinfo {title} {{Phase-space
  holes due to electron and ion beams accelerated by a current-driven potential
  ramp}},}\ }\href@noop {} {\bibfield  {journal} {\bibinfo  {journal}
  {Nonlinear Processes in Geophysics}\ }\textbf {\bibinfo {volume} {10}},\
  \bibinfo {pages} {37--44} (\bibinfo {year} {2003})}\BibitemShut {NoStop}%
\bibitem [{\citenamefont {Wang}\ \emph {et~al.}(2021)\citenamefont {Wang},
  \citenamefont {Vasko}, \citenamefont {Mozer}, \citenamefont {Bale},
  \citenamefont {Kuzichev}, \citenamefont {Artemyev}, \citenamefont
  {Steinvall}, \citenamefont {Ergun}, \citenamefont {Giles}, \citenamefont
  {Khotyaintsev}, \citenamefont {Lindqvist}, \citenamefont {Russell},\ and\
  \citenamefont {Strangeway}}]{Wang2021}%
  \BibitemOpen
  \bibfield  {author} {\bibinfo {author} {\bibfnamefont {R.}~\bibnamefont
  {Wang}}, \bibinfo {author} {\bibfnamefont {I.~Y.}\ \bibnamefont {Vasko}},
  \bibinfo {author} {\bibfnamefont {F.~S.}\ \bibnamefont {Mozer}}, \bibinfo
  {author} {\bibfnamefont {S.~D.}\ \bibnamefont {Bale}}, \bibinfo {author}
  {\bibfnamefont {I.~V.}\ \bibnamefont {Kuzichev}}, \bibinfo {author}
  {\bibfnamefont {A.~V.}\ \bibnamefont {Artemyev}}, \bibinfo {author}
  {\bibfnamefont {K.}~\bibnamefont {Steinvall}}, \bibinfo {author}
  {\bibfnamefont {R.}~\bibnamefont {Ergun}}, \bibinfo {author} {\bibfnamefont
  {B.}~\bibnamefont {Giles}}, \bibinfo {author} {\bibfnamefont
  {Y.}~\bibnamefont {Khotyaintsev}}, \bibinfo {author} {\bibfnamefont {P.-A.}\
  \bibnamefont {Lindqvist}}, \bibinfo {author} {\bibfnamefont {C.~T.}\
  \bibnamefont {Russell}},\ and\ \bibinfo {author} {\bibfnamefont
  {R.}~\bibnamefont {Strangeway}},\ }\bibfield  {title} {\enquote {\bibinfo
  {title} {Electrostatic solitary waves in the earth{\textquotesingle}s bow
  shock: Nature, properties, lifetimes, and origin},}\ }\href
  {https://doi.org/10.1029/2021ja029357} {\bibfield  {journal} {\bibinfo
  {journal} {Journal of Geophysical Research: Space Physics}\ }\textbf
  {\bibinfo {volume} {126}},\ \bibinfo {pages} {e2021JA029357} (\bibinfo {year}
  {2021})}\BibitemShut {NoStop}%
\bibitem [{\citenamefont {Wang}\ \emph {et~al.}(2022)\citenamefont {Wang},
  \citenamefont {Vasko}, \citenamefont {Artemyev}, \citenamefont {Holley},
  \citenamefont {Kamaletdinov}, \citenamefont {Lotekar},\ and\ \citenamefont
  {Mozer}}]{Wang2022}%
  \BibitemOpen
  \bibfield  {author} {\bibinfo {author} {\bibfnamefont {R.}~\bibnamefont
  {Wang}}, \bibinfo {author} {\bibfnamefont {I.~Y.}\ \bibnamefont {Vasko}},
  \bibinfo {author} {\bibfnamefont {A.~V.}\ \bibnamefont {Artemyev}}, \bibinfo
  {author} {\bibfnamefont {L.~C.}\ \bibnamefont {Holley}}, \bibinfo {author}
  {\bibfnamefont {S.~R.}\ \bibnamefont {Kamaletdinov}}, \bibinfo {author}
  {\bibfnamefont {A.}~\bibnamefont {Lotekar}},\ and\ \bibinfo {author}
  {\bibfnamefont {F.~S.}\ \bibnamefont {Mozer}},\ }\bibfield  {title} {\enquote
  {\bibinfo {title} {Multisatellite observations of ion holes in the
  earth{\textquotesingle}s plasma sheet},}\ }\href
  {https://doi.org/10.1029/2022gl097919} {\bibfield  {journal} {\bibinfo
  {journal} {Geophysical Research Letters}\ }\textbf {\bibinfo {volume} {49}},\
  \bibinfo {pages} {e2022GL097919} (\bibinfo {year} {2022})}\BibitemShut
  {NoStop}%
\bibitem [{\citenamefont {Mozer}\ \emph {et~al.}(2021)\citenamefont {Mozer},
  \citenamefont {Bonnell}, \citenamefont {Hanson}, \citenamefont {Gasque},\
  and\ \citenamefont {Vasko}}]{Mozer2021}%
  \BibitemOpen
  \bibfield  {author} {\bibinfo {author} {\bibfnamefont {F.~S.}\ \bibnamefont
  {Mozer}}, \bibinfo {author} {\bibfnamefont {J.~W.}\ \bibnamefont {Bonnell}},
  \bibinfo {author} {\bibfnamefont {E.~L.~M.}\ \bibnamefont {Hanson}}, \bibinfo
  {author} {\bibfnamefont {L.~C.}\ \bibnamefont {Gasque}},\ and\ \bibinfo
  {author} {\bibfnamefont {I.~Y.}\ \bibnamefont {Vasko}},\ }\bibfield  {title}
  {\enquote {\bibinfo {title} {Nonlinear ion-acoustic waves, ion holes, and
  electron holes in the near-sun solar wind},}\ }\href
  {https://doi.org/10.3847/1538-4357/abed52} {\bibfield  {journal} {\bibinfo
  {journal} {The Astrophysical Journal}\ }\textbf {\bibinfo {volume} {911}},\
  \bibinfo {pages} {89} (\bibinfo {year} {2021})}\BibitemShut {NoStop}%
\bibitem [{\citenamefont {Bounds}\ \emph {et~al.}(1999)\citenamefont {Bounds},
  \citenamefont {Pfaff}, \citenamefont {Knowlton}, \citenamefont {Mozer},
  \citenamefont {Temerin},\ and\ \citenamefont {Kletzing}}]{Bounds1999}%
  \BibitemOpen
  \bibfield  {author} {\bibinfo {author} {\bibfnamefont {S.~R.}\ \bibnamefont
  {Bounds}}, \bibinfo {author} {\bibfnamefont {R.~F.}\ \bibnamefont {Pfaff}},
  \bibinfo {author} {\bibfnamefont {S.~F.}\ \bibnamefont {Knowlton}}, \bibinfo
  {author} {\bibfnamefont {F.~S.}\ \bibnamefont {Mozer}}, \bibinfo {author}
  {\bibfnamefont {M.~A.}\ \bibnamefont {Temerin}},\ and\ \bibinfo {author}
  {\bibfnamefont {C.~A.}\ \bibnamefont {Kletzing}},\ }\bibfield  {title}
  {\enquote {\bibinfo {title} {Solitary potential structures associated with
  ion and electron beams near 1re altitude},}\ }\href
  {https://doi.org/10.1029/1999ja900284} {\bibfield  {journal} {\bibinfo
  {journal} {Journal of Geophysical Research: Space Physics}\ }\textbf
  {\bibinfo {volume} {104}},\ \bibinfo {pages} {28709--28717} (\bibinfo {year}
  {1999})}\BibitemShut {NoStop}%
\bibitem [{\citenamefont {Hutchinson}\ and\ \citenamefont
  {Zhou}(2016)}]{Hutchinson2016}%
  \BibitemOpen
  \bibfield  {author} {\bibinfo {author} {\bibfnamefont {I.~H.}\ \bibnamefont
  {Hutchinson}}\ and\ \bibinfo {author} {\bibfnamefont {C.}~\bibnamefont
  {Zhou}},\ }\bibfield  {title} {\enquote {\bibinfo {title} {{Plasma electron
  hole kinematics. I. Momentum conservation}},}\ }\href
  {https://doi.org/10.1063/1.4959870} {\bibfield  {journal} {\bibinfo
  {journal} {Physics of Plasmas}\ }\textbf {\bibinfo {volume} {23}},\ \bibinfo
  {pages} {82101} (\bibinfo {year} {2016})}\BibitemShut {NoStop}%
\bibitem [{\citenamefont {Hutchinson}(2017)}]{Hutchinson2017}%
  \BibitemOpen
  \bibfield  {author} {\bibinfo {author} {\bibfnamefont {I.~H.}\ \bibnamefont
  {Hutchinson}},\ }\bibfield  {title} {\enquote {\bibinfo {title} {{Electron
  holes in phase space: What they are and why they matter}},}\ }\href
  {https://doi.org/10.1063/1.4976854} {\bibfield  {journal} {\bibinfo
  {journal} {Physics of Plasmas}\ }\textbf {\bibinfo {volume} {24}},\ \bibinfo
  {pages} {055601} (\bibinfo {year} {2017})}\BibitemShut {NoStop}%
\bibitem [{\citenamefont {Muschietti}\ and\ \citenamefont
  {Roth}(2008)}]{Muschietti2008}%
  \BibitemOpen
  \bibfield  {author} {\bibinfo {author} {\bibfnamefont {L.}~\bibnamefont
  {Muschietti}}\ and\ \bibinfo {author} {\bibfnamefont {I.}~\bibnamefont
  {Roth}},\ }\bibfield  {title} {\enquote {\bibinfo {title} {Ion two-stream
  instabilities in the auroral acceleration zone},}\ }\href
  {https://doi.org/10.1029/2007ja013005} {\bibfield  {journal} {\bibinfo
  {journal} {Journal of Geophysical Research: Space Physics}\ }\textbf
  {\bibinfo {volume} {113}},\ \bibinfo {pages} {A08201} (\bibinfo {year}
  {2008})}\BibitemShut {NoStop}%
\bibitem [{\citenamefont {Hutchinson}(2021{\natexlab{a}})}]{Hutchinson2021d}%
  \BibitemOpen
  \bibfield  {author} {\bibinfo {author} {\bibfnamefont {I.~H.}\ \bibnamefont
  {Hutchinson}},\ }\bibfield  {title} {\enquote {\bibinfo {title} {Asymmetric
  one-dimensional slow electron holes},}\ }\href
  {https://doi.org/10.1103/PhysRevE.104.055207} {\bibfield  {journal} {\bibinfo
   {journal} {Phys. Rev. E}\ }\textbf {\bibinfo {volume} {104}},\ \bibinfo
  {pages} {055207} (\bibinfo {year} {2021}{\natexlab{a}})}\BibitemShut
  {NoStop}%
\bibitem [{\citenamefont {Hutchinson}(2021{\natexlab{b}})}]{Hutchinson2021c}%
  \BibitemOpen
  \bibfield  {author} {\bibinfo {author} {\bibfnamefont {I.~H.}\ \bibnamefont
  {Hutchinson}},\ }\bibfield  {title} {\enquote {\bibinfo {title} {How can slow
  plasma electron holes exist?}}\ }\href
  {https://doi.org/10.1103/PhysRevE.104.015208} {\bibfield  {journal} {\bibinfo
   {journal} {Phys. Rev. E}\ }\textbf {\bibinfo {volume} {104}},\ \bibinfo
  {pages} {015208} (\bibinfo {year} {2021}{\natexlab{b}})},\ \Eprint
  {https://arxiv.org/abs/http://arxiv.org/abs/2104.13800}
  {http://arxiv.org/abs/2104.13800} \BibitemShut {NoStop}%
\bibitem [{Note1()}]{Note1}%
  \BibitemOpen
  \bibinfo {note} {The Faddeeva function is related to the Plasma Dispersion
  function via $Z(z)=i\protect \sqrt {\pi } w(z)$ and to the Dawson integral
  function $F(z)$ by $w(iz)=2F(z)$. Its derivative is $w'(z)=2i/\protect \sqrt
  {\pi }-2zw(z)$; so $-Z'(z)/2=1+i\protect \sqrt {\pi }zw(z)$.}\BibitemShut
  {Stop}%
\bibitem [{\citenamefont {Chen}, \citenamefont {Thouless},\ and\ \citenamefont
  {Tang}(2004)}]{Chen2004}%
  \BibitemOpen
  \bibfield  {author} {\bibinfo {author} {\bibfnamefont {L.-J.}\ \bibnamefont
  {Chen}}, \bibinfo {author} {\bibfnamefont {D.}~\bibnamefont {Thouless}},\
  and\ \bibinfo {author} {\bibfnamefont {J.-M.}\ \bibnamefont {Tang}},\
  }\bibfield  {title} {\enquote {\bibinfo {title}
  {{Bernstein–Greene–Kruskal solitary waves in three-dimensional magnetized
  plasma}},}\ }\href {https://doi.org/10.1103/PhysRevE.69.055401} {\bibfield
  {journal} {\bibinfo  {journal} {Physical Review E}\ }\textbf {\bibinfo
  {volume} {69}},\ \bibinfo {pages} {55401} (\bibinfo {year}
  {2004})}\BibitemShut {NoStop}%
\bibitem [{Note2()}]{Note2}%
  \BibitemOpen
  \bibinfo {note} {Those authors used a Gaussian potential shape so their
  $\protect \tilde f_\varphi $ is different and does not satisfy eq.\ (\ref
  {eq:shielding})}\BibitemShut {NoStop}%
\bibitem [{Note3()}]{Note3}%
  \BibitemOpen
  \bibinfo {note} {Schamel et al proceed instead by the differential approach,
  specifying a negative temperature Maxwellian for trapped particles. They find
  by expansion at small $\psi $ that $\varphi =\psi {\protect \,\protect \rm
  sech}^4(z/4\lambda )$, with $\lambda ^{-2}=16b\protect \sqrt {{\psi }}/15$
  where $b$ depends on the trapped temperature and $\protect \bar v_a$, and
  that (\ref {eq:modshield}) $1/\lambda ^2-1/\theta +{1 \over 2} Z'_r(\protect
  \bar v_a/\protect \sqrt {2})=0$, which they call the nonlinear dispersion
  relation.}\BibitemShut {Stop}%
\bibitem [{\citenamefont {Schamel}(1979)}]{Schamel1979}%
  \BibitemOpen
  \bibfield  {author} {\bibinfo {author} {\bibfnamefont {H.}~\bibnamefont
  {Schamel}},\ }\bibfield  {title} {\enquote {\bibinfo {title} {{Theory of
  Electron Holes}},}\ }\href {https://doi.org/10.1088/0031-8949/20/3-4/006}
  {\bibfield  {journal} {\bibinfo  {journal} {Physica Scripta}\ }\textbf
  {\bibinfo {volume} {20}},\ \bibinfo {pages} {336--342} (\bibinfo {year}
  {1979})}\BibitemShut {NoStop}%
\bibitem [{\citenamefont {Schamel}\ and\ \citenamefont
  {Bujarbarua}(1980)}]{Schamel1980}%
  \BibitemOpen
  \bibfield  {author} {\bibinfo {author} {\bibfnamefont {H.}~\bibnamefont
  {Schamel}}\ and\ \bibinfo {author} {\bibfnamefont {S.}~\bibnamefont
  {Bujarbarua}},\ }\bibfield  {title} {\enquote {\bibinfo {title} {Solitary
  plasma hole via ion‐vortex distribution},}\ }\href
  {https://doi.org/.org/10.1063/1.4894115} {\bibfield  {journal} {\bibinfo
  {journal} {Physics of Fluids}\ }\textbf {\bibinfo {volume} {23}},\ \bibinfo
  {pages} {2498--2499} (\bibinfo {year} {1980})}\BibitemShut {NoStop}%
\bibitem [{\citenamefont {Hudson}\ \emph {et~al.}(1983)\citenamefont {Hudson},
  \citenamefont {Lotko}, \citenamefont {Roth},\ and\ \citenamefont
  {Witt}}]{Hudson1983}%
  \BibitemOpen
  \bibfield  {author} {\bibinfo {author} {\bibfnamefont {M.}~\bibnamefont
  {Hudson}}, \bibinfo {author} {\bibfnamefont {W.}~\bibnamefont {Lotko}},
  \bibinfo {author} {\bibfnamefont {I.}~\bibnamefont {Roth}},\ and\ \bibinfo
  {author} {\bibfnamefont {E.}~\bibnamefont {Witt}},\ }\bibfield  {title}
  {\enquote {\bibinfo {title} {Solitary waves and double layers on auroral
  field lines},}\ }\href@noop {} {\bibfield  {journal} {\bibinfo  {journal} {J
  Geophysical Research A}\ }\textbf {\bibinfo {volume} {88}},\ \bibinfo {pages}
  {916--926} (\bibinfo {year} {1983})}\BibitemShut {NoStop}%
\bibitem [{\citenamefont {Buchanan}\ and\ \citenamefont
  {Doming}(1993)}]{Buchanan1993}%
  \BibitemOpen
  \bibfield  {author} {\bibinfo {author} {\bibfnamefont {M.}~\bibnamefont
  {Buchanan}}\ and\ \bibinfo {author} {\bibfnamefont {J.~J.}\ \bibnamefont
  {Doming}},\ }\bibfield  {title} {\enquote {\bibinfo {title} {Nonlinear waves
  in collisionless plasmas},}\ }\href@noop {} {\bibfield  {journal} {\bibinfo
  {journal} {Physics Letters A}\ }\textbf {\bibinfo {volume} {179}},\ \bibinfo
  {pages} {306--310} (\bibinfo {year} {1993})}\BibitemShut {NoStop}%
\bibitem [{\citenamefont {Grie{\ss}meier}\ and\ \citenamefont
  {Schamel}(2002)}]{Griessmeier2002}%
  \BibitemOpen
  \bibfield  {author} {\bibinfo {author} {\bibfnamefont {J.-M.}\ \bibnamefont
  {Grie{\ss}meier}}\ and\ \bibinfo {author} {\bibfnamefont {H.}~\bibnamefont
  {Schamel}},\ }\bibfield  {title} {\enquote {\bibinfo {title} {Solitary holes
  of negative energy and their possible role in the nonlinear destabilization
  of plasmas},}\ }\href {https://doi.org/10.1063/1.1477450} {\bibfield
  {journal} {\bibinfo  {journal} {Physics of Plasmas}\ }\textbf {\bibinfo
  {volume} {9}},\ \bibinfo {pages} {2462--2465} (\bibinfo {year}
  {2002})}\BibitemShut {NoStop}%
\bibitem [{\citenamefont {Eliasson}, \citenamefont {Shukla},\ and\
  \citenamefont {Dieckmann}(2006)}]{Eliasson2006a}%
  \BibitemOpen
  \bibfield  {author} {\bibinfo {author} {\bibfnamefont {B.}~\bibnamefont
  {Eliasson}}, \bibinfo {author} {\bibfnamefont {P.~K.}\ \bibnamefont
  {Shukla}},\ and\ \bibinfo {author} {\bibfnamefont {M.~E.}\ \bibnamefont
  {Dieckmann}},\ }\bibfield  {title} {\enquote {\bibinfo {title} {Theoretical
  and simulation studies of relativistic ion holes in astrophysical plasmas},}\
  }\href {https://doi.org/10.1088/1367-2630/8/4/055} {\bibfield  {journal}
  {\bibinfo  {journal} {New Journal of Physics}\ }\textbf {\bibinfo {volume}
  {8}},\ \bibinfo {pages} {55--55} (\bibinfo {year} {2006})}\BibitemShut
  {NoStop}%
\bibitem [{\citenamefont {Schamel}, \citenamefont {Das},\ and\ \citenamefont
  {Borah}(2018)}]{Schamel2018}%
  \BibitemOpen
  \bibfield  {author} {\bibinfo {author} {\bibfnamefont {H.}~\bibnamefont
  {Schamel}}, \bibinfo {author} {\bibfnamefont {N.}~\bibnamefont {Das}},\ and\
  \bibinfo {author} {\bibfnamefont {P.}~\bibnamefont {Borah}},\ }\bibfield
  {title} {\enquote {\bibinfo {title} {The privileged spectrum of cnoidal ion
  holes and its extension by imperfect ion trapping},}\ }\href
  {https://doi.org/10.1016/j.physleta.2017.11.004} {\bibfield  {journal}
  {\bibinfo  {journal} {Physics Letters A}\ }\textbf {\bibinfo {volume}
  {382}},\ \bibinfo {pages} {168--174} (\bibinfo {year} {2018})}\BibitemShut
  {NoStop}%
\bibitem [{Note4()}]{Note4}%
  \BibitemOpen
  \bibinfo {note} {A hole requires the inequality to be the opposite of the
  criterion for existence of what {\protect \citet {Stix1962}} [section 9.14,
  equation 71] calls the ``zero-damped ion acoustic wave'', because the
  sinusoidal shape of a wave makes ${1\over \lambda ^2} \equiv {1\over \phi
  }{d^2\phi \over dz^2}$ negative, not positive.}\BibitemShut {Stop}%
\bibitem [{\citenamefont {Hutchinson}(2021{\natexlab{c}})}]{Hutchinson2021a}%
  \BibitemOpen
  \bibfield  {author} {\bibinfo {author} {\bibfnamefont {I.~H.}\ \bibnamefont
  {Hutchinson}},\ }\bibfield  {title} {\enquote {\bibinfo {title} {Synthetic
  multidimensional plasma electron hole equilibria},}\ }\href
  {http://arxiv.org/abs/2101.09275} {\bibfield  {journal} {\bibinfo  {journal}
  {Physics of Plasmas}\ }\textbf {\bibinfo {volume} {26}},\ \bibinfo {pages}
  {062036} (\bibinfo {year} {2021}{\natexlab{c}})}\BibitemShut {NoStop}%
\bibitem [{\citenamefont {Dupree}(1983)}]{Dupree1983}%
  \BibitemOpen
  \bibfield  {author} {\bibinfo {author} {\bibfnamefont {T.~H.}\ \bibnamefont
  {Dupree}},\ }\bibfield  {title} {\enquote {\bibinfo {title} {{Growth of
  phase-space density holes}},}\ }\href {https://doi.org/10.1063/1.864430}
  {\bibfield  {journal} {\bibinfo  {journal} {Physics of Fluids}\ }\textbf
  {\bibinfo {volume} {26}},\ \bibinfo {pages} {2460} (\bibinfo {year}
  {1983})}\BibitemShut {NoStop}%
\bibitem [{\citenamefont {Eliasson}\ and\ \citenamefont
  {Shukla}(2004)}]{Eliasson2004}%
  \BibitemOpen
  \bibfield  {author} {\bibinfo {author} {\bibfnamefont {B.}~\bibnamefont
  {Eliasson}}\ and\ \bibinfo {author} {\bibfnamefont {P.~K.}\ \bibnamefont
  {Shukla}},\ }\bibfield  {title} {\enquote {\bibinfo {title} {{Dynamics of
  electron holes in an electron-oxygen-ion plasma.}}}\ }\href
  {https://doi.org/10.1103/PhysRevLett.93.045001} {\bibfield  {journal}
  {\bibinfo  {journal} {Physical Review Letters}\ }\textbf {\bibinfo {volume}
  {93}},\ \bibinfo {pages} {45001} (\bibinfo {year} {2004})}\BibitemShut
  {NoStop}%
\bibitem [{\citenamefont {Eliasson}\ and\ \citenamefont
  {Shukla}(2006)}]{Eliasson2006}%
  \BibitemOpen
  \bibfield  {author} {\bibinfo {author} {\bibfnamefont {B.}~\bibnamefont
  {Eliasson}}\ and\ \bibinfo {author} {\bibfnamefont {P.~K.}\ \bibnamefont
  {Shukla}},\ }\bibfield  {title} {\enquote {\bibinfo {title} {{Formation and
  dynamics of coherent structures involving phase-space vortices in
  plasmas}},}\ }\href {https://doi.org/10.1016/j.physrep.2005.10.003}
  {\bibfield  {journal} {\bibinfo  {journal} {Physics Reports}\ }\textbf
  {\bibinfo {volume} {422}},\ \bibinfo {pages} {225--290} (\bibinfo {year}
  {2006})}\BibitemShut {NoStop}%
\bibitem [{\citenamefont {Zhou}\ and\ \citenamefont
  {Hutchinson}(2016)}]{Zhou2016}%
  \BibitemOpen
  \bibfield  {author} {\bibinfo {author} {\bibfnamefont {C.}~\bibnamefont
  {Zhou}}\ and\ \bibinfo {author} {\bibfnamefont {I.~H.}\ \bibnamefont
  {Hutchinson}},\ }\bibfield  {title} {\enquote {\bibinfo {title} {{Plasma
  electron hole kinematics. II. Hole tracking Particle-In-Cell simulation}},}\
  }\href {https://doi.org/10.1063/1.4959871} {\bibfield  {journal} {\bibinfo
  {journal} {Physics of Plasmas}\ }\textbf {\bibinfo {volume} {23}},\ \bibinfo
  {pages} {82102} (\bibinfo {year} {2016})}\BibitemShut {NoStop}%
\bibitem [{\citenamefont {Kamaletdinov}\ \emph {et~al.}(2021)\citenamefont
  {Kamaletdinov}, \citenamefont {Hutchinson}, \citenamefont {Vasko},
  \citenamefont {Artemyev}, \citenamefont {Lotekar},\ and\ \citenamefont
  {Mozer}}]{Kamaletdinov2021}%
  \BibitemOpen
  \bibfield  {author} {\bibinfo {author} {\bibfnamefont {S.~R.}\ \bibnamefont
  {Kamaletdinov}}, \bibinfo {author} {\bibfnamefont {I.~H.}\ \bibnamefont
  {Hutchinson}}, \bibinfo {author} {\bibfnamefont {I.~Y.}\ \bibnamefont
  {Vasko}}, \bibinfo {author} {\bibfnamefont {A.}~\bibnamefont {Artemyev}},
  \bibinfo {author} {\bibfnamefont {A.}~\bibnamefont {Lotekar}},\ and\ \bibinfo
  {author} {\bibfnamefont {F.}~\bibnamefont {Mozer}},\ }\bibfield  {title}
  {\enquote {\bibinfo {title} {Spacecraft observations and theoretical
  understanding of slow electron holes},}\ }\href
  {https://doi.org/10.1103/PhysRevLett.127.165101} {\bibfield  {journal}
  {\bibinfo  {journal} {Phys. Rev. Lett.}\ }\textbf {\bibinfo {volume} {127}},\
  \bibinfo {pages} {165101} (\bibinfo {year} {2021})}\BibitemShut {NoStop}%
\bibitem [{\citenamefont {Chen}\ and\ \citenamefont
  {Hutchinson}(2023)}]{Chen2023}%
  \BibitemOpen
  \bibfield  {author} {\bibinfo {author} {\bibfnamefont {X.}~\bibnamefont
  {Chen}}\ and\ \bibinfo {author} {\bibfnamefont {I.~H.}\ \bibnamefont
  {Hutchinson}},\ }\bibfield  {title} {\enquote {\bibinfo {title} {Multimode
  theory of electron hole transverse instability},}\ }\href@noop {} {\bibfield
  {journal} {\bibinfo  {journal} {Journal of Plasma Physics}\ ,\ \bibinfo
  {pages} {to appear}} (\bibinfo {year} {2023})}\BibitemShut {NoStop}%
\bibitem [{\citenamefont {Hutchinson}(2022)}]{Hutchinson2022}%
  \BibitemOpen
  \bibfield  {author} {\bibinfo {author} {\bibfnamefont {I.}~\bibnamefont
  {Hutchinson}},\ }\bibfield  {title} {\enquote {\bibinfo {title}
  {Overstability of plasma slow electron holes},}\ }\href
  {https://doi.org/10.1017/s0022377822000149} {\bibfield  {journal} {\bibinfo
  {journal} {Journal of Plasma Physics}\ }\textbf {\bibinfo {volume} {88}},\
  \bibinfo {pages} {555880101} (\bibinfo {year} {2022})}\BibitemShut {NoStop}%
\bibitem [{\citenamefont {Hutchinson}(2018{\natexlab{a}})}]{Hutchinson2018}%
  \BibitemOpen
  \bibfield  {author} {\bibinfo {author} {\bibfnamefont {I.~H.}\ \bibnamefont
  {Hutchinson}},\ }\bibfield  {title} {\enquote {\bibinfo {title} {{Kinematic
  Mechanism of Plasma Electron Hole Transverse Instability}},}\ }\href
  {https://doi.org/10.1103/PhysRevLett.120.205101} {\bibfield  {journal}
  {\bibinfo  {journal} {Physical Review Letters}\ }\textbf {\bibinfo {volume}
  {120}},\ \bibinfo {pages} {205101} (\bibinfo {year}
  {2018}{\natexlab{a}})}\BibitemShut {NoStop}%
\bibitem [{\citenamefont {Hutchinson}(2018{\natexlab{b}})}]{Hutchinson2018a}%
  \BibitemOpen
  \bibfield  {author} {\bibinfo {author} {\bibfnamefont {I.~H.}\ \bibnamefont
  {Hutchinson}},\ }\bibfield  {title} {\enquote {\bibinfo {title} {{Transverse
  instability of electron phase-space holes in multi-dimensional Maxwellian
  plasmas}},}\ }\href {https://doi.org/10.1017/S0022377818000909} {\bibfield
  {journal} {\bibinfo  {journal} {Journal of Plasma Physics}\ }\textbf
  {\bibinfo {volume} {84}},\ \bibinfo {pages} {905840411} (\bibinfo {year}
  {2018}{\natexlab{b}})},\ \Eprint {https://arxiv.org/abs/1804.08594}
  {arXiv:1804.08594} \BibitemShut {NoStop}%
\bibitem [{\citenamefont {Haas}(2021)}]{Haas2021}%
  \BibitemOpen
  \bibfield  {author} {\bibinfo {author} {\bibfnamefont {F.}~\bibnamefont
  {Haas}},\ }\bibfield  {title} {\enquote {\bibinfo {title} {Electron holes in
  a kappa distribution background with singularities},}\ }\href@noop {}
  {\bibfield  {journal} {\bibinfo  {journal} {Physics of Plasma}\ }\textbf
  {\bibinfo {volume} {28}},\ \bibinfo {pages} {072110} (\bibinfo {year}
  {2021})}\BibitemShut {NoStop}%
\bibitem [{\citenamefont {Lesur}, \citenamefont {Diamond},\ and\ \citenamefont
  {Kosuga}(2014)}]{Lesur2014}%
  \BibitemOpen
  \bibfield  {author} {\bibinfo {author} {\bibfnamefont {M.}~\bibnamefont
  {Lesur}}, \bibinfo {author} {\bibfnamefont {P.~H.}\ \bibnamefont {Diamond}},\
  and\ \bibinfo {author} {\bibfnamefont {Y.}~\bibnamefont {Kosuga}},\
  }\bibfield  {title} {\enquote {\bibinfo {title} {{Nonlinear current-driven
  ion-acoustic instability driven by phase-space structures}},}\ }\href
  {https://doi.org/10.1088/0741-3335/56/7/075005} {\bibfield  {journal}
  {\bibinfo  {journal} {Plasma Physics and Controlled Fusion}\ }\textbf
  {\bibinfo {volume} {56}},\ \bibinfo {pages} {75005} (\bibinfo {year}
  {2014})}\BibitemShut {NoStop}%
\bibitem [{\citenamefont {Hutchinson}(2020)}]{Hutchinson2020}%
  \BibitemOpen
  \bibfield  {author} {\bibinfo {author} {\bibfnamefont {I.~H.}\ \bibnamefont
  {Hutchinson}},\ }\bibfield  {title} {\enquote {\bibinfo {title} {Particle
  trapping in axisymmetric electron holes},}\ }\href
  {https://doi.org/10.1029/2020JA028093} {\bibfield  {journal} {\bibinfo
  {journal} {Journal of Geophysical Research: Space Physics}\ }\textbf
  {\bibinfo {volume} {125}} (\bibinfo {year} {2020}),\ 10.1029/2020JA028093},\
  \bibinfo {note} {e2020JA028093 10.1029/2020JA028093},\ \Eprint
  {https://arxiv.org/abs/https://agupubs.onlinelibrary.wiley.com/doi/pdf/10.1029/2020JA028093}
  {https://agupubs.onlinelibrary.wiley.com/doi/pdf/10.1029/2020JA028093}
  \BibitemShut {NoStop}%
\bibitem [{\citenamefont {Bohm}\ and\ \citenamefont {Gross}(1949)}]{Bohm1949}%
  \BibitemOpen
  \bibfield  {author} {\bibinfo {author} {\bibfnamefont {D.}~\bibnamefont
  {Bohm}}\ and\ \bibinfo {author} {\bibfnamefont {E.~P.}\ \bibnamefont
  {Gross}},\ }\bibfield  {title} {\enquote {\bibinfo {title} {Theory of plasma
  oscillations. a. origin of medium-like behavior},}\ }\href@noop {} {\bibfield
   {journal} {\bibinfo  {journal} {Physical Review}\ }\textbf {\bibinfo
  {volume} {75}},\ \bibinfo {pages} {1851--1864} (\bibinfo {year}
  {1949})}\BibitemShut {NoStop}%
\bibitem [{\citenamefont {Stix}(1962)}]{Stix1962}%
  \BibitemOpen
  \bibfield  {author} {\bibinfo {author} {\bibfnamefont {T.~H.}\ \bibnamefont
  {Stix}},\ }\href@noop {} {\emph {\bibinfo {title} {The theory of plasma
  Waves}}}\ (\bibinfo  {publisher} {McGraw-Hill, New York},\ \bibinfo {year}
  {1962})\BibitemShut {NoStop}%
\end{thebibliography}%

\end{document}